%% file: AA18731.tex
\documentclass[a4paper,traditabstract,longauth]{aa}  
\usepackage{graphicx}
\usepackage{txfonts}
\usepackage[breaklinks, colorlinks, citecolor=blue]{hyperref}
\usepackage{url}
\usepackage{natbib}
\usepackage{color}
\usepackage{amssymb}
\usepackage{booktabs}
\usepackage{subfig}

\def\ergscm{\rm ergs\,s^{-1}\,cm^{-2}}

\def\msol{{M$_{\odot}$}}
\def\xmm{XMM-{\it Newton} }

\def\xmm{{\it XMM-Newton}}

\def\planck{{\it Planck}}
\def\rosat{{\it ROSAT}}
\def\rass{{\rm RASS}}

\def \epic {\hbox{\sc EPIC}} 
\def \mos {\hbox{\sc EMOS}} 
\def \pn {\hbox{\sc EPN}} 
\def \mekal {\hbox{\sc mekal}}

\newfont{\gwpfont}{cmssq8 scaled 1000}
\newcommand{\rexcess}{{\gwpfont REXCESS}}
\newcommand{\excpres}{{\gwpfont EXCPRES}}
\newcommand{\reflex}{{REFLEX}}
\newcommand{\noras}{{NORAS}}
\newcommand{\bcs}{{BCS}}
\newcommand{\ebcs}{{eBCS}}
\newcommand{\macs}{{MACS}}
\def\Mv{M_{500}}
\def\Rv{R_{500}}
\def\Mgv{M_{\rm g,500}}

\def\YX {Y_{\rm X}}
\def\TX {T_{\rm X}}
\def\kT {{\rm k}T}
\def\YSZ {Y_{\rm SZ}}
\def\YSZ {Y_{500}}

\def\kT {{\rm k}T}
\def\Mv {M_{\rm 500}}
\def \Rv {R_{500}}
\def\keV {\rm keV}
\def\Yv {Y_{500}}
\def\LX {L_{500,[0.1-2.4]\,\keV}}

\def\MYX {$M_{500}$--$Y_{\rm X}$}

\def\Lxz{$L_{\rm X}$--$z$}

\def\YSZYX {$\YSZ$--$\YX$}
\def\YXYSZ{$\YX$--$\YSZ$}

\def\LXYSZ {$L_{\rm X}$--$\YSZ$}

\def\msol {{\rm M_{\odot}}}

\def\lesssim{\mathrel{\hbox{\rlap{\hbox{\lower4pt\hbox{$\sim$}}}\hbox{$<$}}}}
\def\gtrsim{\mathrel{\hbox{\rlap{\hbox{\lower4pt\hbox{$\sim$}}}\hbox{$>$}}}}

\newcommand{\propsim}{\lower 3pt \hbox{$\, \buildrel {\textstyle
       \propto}\over {\textstyle \sim}\,$}}

\begin{document}
%

\input{PIP_22a_authors_and_institutes.tex}

\title{ \textit{Planck} Intermediate  Results. I.  Further validation  of new \textit{Planck} clusters  with \textit{XMM-Newton}}
 \date{Received  December 23; Accepted  April 15}
  \abstract
 { We present further results from the ongoing \xmm\ validation follow-up of \planck\ cluster candidates, detailing X-ray observations of eleven candidates detected at a signal-to-noise ratio of $4.5 < {\rm S/N} < 5.3$ in the same 10-month survey maps used in the construction of the Early SZ sample. The sample was selected in order to test internal SZ quality flags, and the pertinence of these flags is discussed in light of the validation results.  Ten of the candidates are found to be {\it bona fide} clusters lying below the \rass\ flux limit.  Redshift estimates are available for all confirmed systems via X-ray Fe-line spectroscopy. They lie in the redshift range $0.19 < z <0.94$, demonstrating \planck's  capability to detect clusters up to high $z$. 
  The X-ray properties of the new clusters appear to be similar to previous new detections by \planck\  at lower $z$ and higher SZ flux: the majority are X-ray underluminous for their mass, estimated using $Y_{\rm X}$ as mass proxy, and many have a disturbed morphology. We find tentative indication for Malmquist bias in the $Y_{\rm SZ}$--$Y_{\rm X}$ relation, with a turnover at $Y_{\rm SZ} \sim 4\times10^{-4}$ arcmin$^2$. 
  We present additional new optical redshift determinations with ENO and ESO  telescopes of candidates previously confirmed with \xmm. The X-ray and optical redshifts for a total of 20 clusters are found to be in excellent agreement. We also show that useful lower limits can be put on cluster redshifts using X-ray data only via the use of the $Y_{\rm X}$ vs. $Y_{\rm SZ}$ and X-ray flux $F_{\rm X}$ vs. $Y_{\rm SZ}$ relations.}
   \keywords{Cosmology: observations $-$  Galaxies: cluster: general $-$ Galaxies: clusters: intracluster medium $-$ Cosmic background radiation, X-rays: galaxies: clusters}

\authorrunning{Planck Collaboration}
\titlerunning{Further confirmation of new \planck\  clusters with \xmm}
  \maketitle

\input{Planck.tex}

\section{Introduction}

The deep potential wells in clusters of galaxies make them unique laboratories in which to study astrophysical processes linked to gas physics, galaxy formation, and feedback. Furthermore, since clusters trace the highest peaks of the matter density field, the properties of the cluster population and their  evolution are a sensitive cosmological probe.

The recent advent of increased sensitivity and survey capability has transformed  galaxy cluster searches via the Sunyaev-Zeldovich (SZ) effect. Such surveys identify objects using the spectral distortion of the cosmic microwave background (CMB) generated through inverse Compton scattering of CMB photons by the hot electrons in the intra-cluster medium \citep{sun72}. Crucially, the total SZ signal is expected to be closely related to the cluster mass \citep[e.g.,][]{das04}, and its surface brightness insensitive to distance. As a result, SZ surveys can potentially provide unbiased cluster samples  over a wide range of redshifts  that are as close as possible to being mass-selected. Such samples are essential for understanding the statistical properties of the cluster population and for its exploitation in cosmological studies. Examples of on-going cluster surveys in the SZ include the Atacama Cosmology Telescope \citep[ACT][]{mar11}, \planck\footnote{\Planck\ (http://www.esa.int/Planck) is a project of the European Space Agency (ESA) with instruments provided by two scientific consortia funded by ESA member states (in particular the lead countries: France and Italy) with contributions from NASA (USA), and telescope reflectors provided in a collaboration between ESA and a scientific consortium led and funded by Denmark.} \citep{planck2011-5.1b} and the South Pole Telescope  \citep[SPT][]{car09a}. 

The \planck\ satellite has been surveying the sky in the microwave band since August 2009 \citep{planck2011-1.1}. Compared to other SZ surveys,  \planck\  has only modest (band-dependent) spatial resolution of 5\arcmin to 30\arcmin\ \citep{planck2011-1.4,planck2011-1.5} but it possesses unique nine-band coverage from 30 to 857\,GHz and, most crucially, it covers an exceptionally large survey volume. Indeed \planck\ is the first all-sky survey capable of blind cluster detections since the \rosat\  All-Sky Survey (\rass, in the X-ray domain). Early \planck\ results on galaxy clusters were recently published in   \citet{planck2011-5.1a,planck2011-5.1b,planck2011-5.1c,planck2011-5.2a,planck2011-5.2b,planck2011-5.2c}. These results include the publication of the high signal--to--noise ratio (${\rm S/N} > 6$) Early SZ (ESZ) cluster sample  \citep{planck2011-5.1a} 

The raw data product of any cluster survey is a list of potential candidates. Such a list is expected to include a fraction of false detections, e.g., for SZ detections, due to fluctuations in the complex microwave astrophysical sky. In the case of \planck, the moderate spatial resolution at SZ frequencies with respect to typical cluster sizes presents a further complication. A \planck\ cluster SZ measurement essentially provides only a position, a total SZ flux, and a coarse size estimate. In addition, the quality of the SZ flux estimates is degraded by the flux-size degeneracy, as discussed in \citet{planck2011-5.1a}. A  follow-up programme is therefore required to scientifically exploit \planck\ candidate data. Such a programme should provide candidate confirmation, which is the final part of the catalogue validation, and a redshift measurement, the prerequisite to any cluster physical parameter estimate. 

In this context, X-ray observations are extremely useful, as has been shown by the results from the initial validation follow-up of \planck\ cluster candidates with \xmm\  \citep{planck2011-5.1b}. These observations were undertaken in Director's Discretionary Time via an agreement between the \xmm\ and \planck\ Project Scientists. A pilot programme observed ten targets to refine the selection criteria for the ESZ cluster sample. A second programme focused on the validation of fifteen high-significance SZ sources (${\rm S/N} > 5$); eleven of the  newly-discovered clusters from this programme are contained in the ESZ sample.  These first observations provided a preview of the X--ray properties of the newly-discovered clusters \citep{planck2011-5.1b}. In particular it was  confirmed that, based on the detection of extended emission, \xmm\  snapshot exposures (10\,ksec) are sufficient for unambiguous discrimination between clusters and false candidates for redshifts at least up to $z = 1.5$. In addition, it was shown that the spurious association of candidates with faint extended sources lying within the \planck\  position uncertainty (which can be up to $5\arcmin$) can be identified via a consistency check between the X-ray and SZ flux. This latter constraint stems from the tight correlation between X-ray and SZ properties, since X-rays probe the same medium as the SZE. In this respect, X-ray validation presents a clear advantage over optical validation for \planck\ candidates.  While  optical observations offer important complementary information on the stellar component of clusters and on mass estimates derived from gravitational lensing of background sources, optical validation is hampered by the relatively large \planck\ source position uncertainty and the large scatter between simple optical observables (such as galaxy numbers) and the mass (or SZ signal), both of which increase the chance of false associations.

A manageable confirmation programme for the compilation of a larger, final, cluster catalogue from the \planck\ survey requires a candidate sample with a high ratio of true clusters to total candidates (i.e., purity). The construction of such a sample relies both on \planck\ internal candidate selection and assessment of the SZ signal quality and also on cross-correlation with ancillary data and catalogues, as described in \citet{planck2011-5.1a}. In the present paper, in which we report on a further eleven \xmm\ observations of \planck\ cluster candidates detected at  $4.5\!<\!{\rm S/N}\!<\!5.3$, we address in more detail the internal quality assessment of cluster candidates in SZ. \xmm\ validation, allowing unambiguous discrimination between clusters and false candidates, is essential for such a study.

X-ray observations can also constrain the redshift of the source through Fe~K line spectroscopy, as demonstrated in \citet{planck2011-5.1b}. Here we also  present new optical redshift determinations for \xmm\ confirmed candidates, which we compare to the X-ray-derived values. We also discuss whether, in the absence of optical follow-up data, a combined X-ray/SZ analysis can improve the $z$ estimate when X-ray data alone are insufficient to unambiguously determine the redshift.

We adopt a $\Lambda$CDM cosmology with $H_0=70\,\kmsMpc$, $\Omega_{\rm M}=0.3$, and $\Omega_\Lambda=0.7$. The factor $E(z)= \sqrt{\Omega_{\rm M} (1+z)^3+\Omega_\Lambda}$ is the ratio of the Hubble constant at redshift $z$ to its present-day value. The quantities $\Mv$ and $\Rv$ are the total mass and radius corresponding to a total density contrast $\delta=500$, as compared to $\rho_{\rm c}(z)$, the critical density of the Universe at the cluster redshift;  $\Mv = (4\pi/3)\,500\,\rho_c(z)\,\Rv^3$. The SZ flux is characterised by $\YSZ$, where $\YSZ\,D_{\rm A}^2$ is the spherically integrated Compton parameter within $\Rv$, and $D_{\rm A}$ is the angular-diameter distance to the cluster.

%________________________________________________________________
%% Figure: 
%%
\begin{figure*}[t]
\centering
\begin{minipage}[t]{0.85\textwidth}
\resizebox{\hsize}{!} {
\includegraphics[scale=1.,angle=0,keepaspectratio]{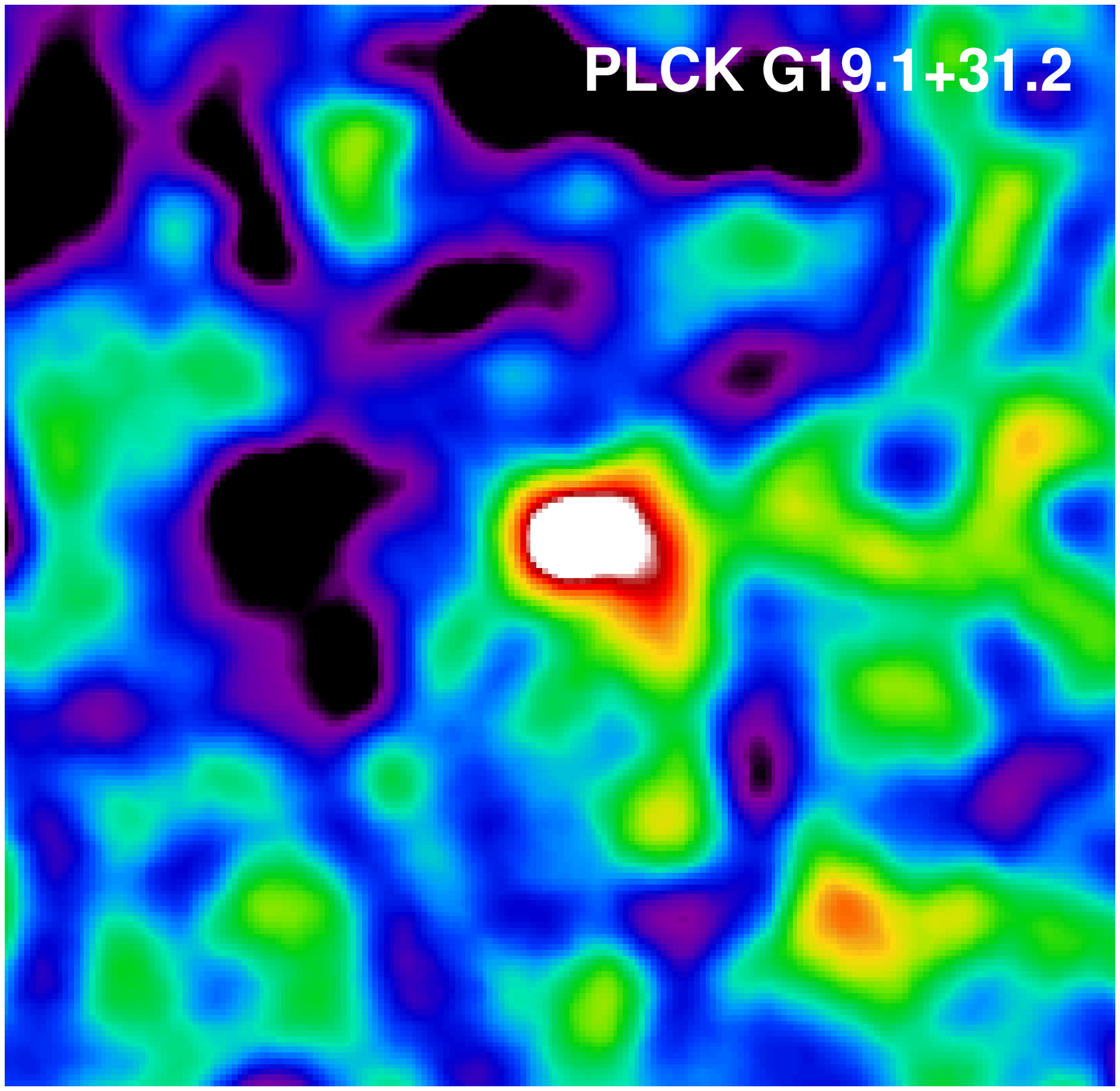}%
\hspace{5mm}
\includegraphics[scale=1.,angle=0,keepaspectratio]{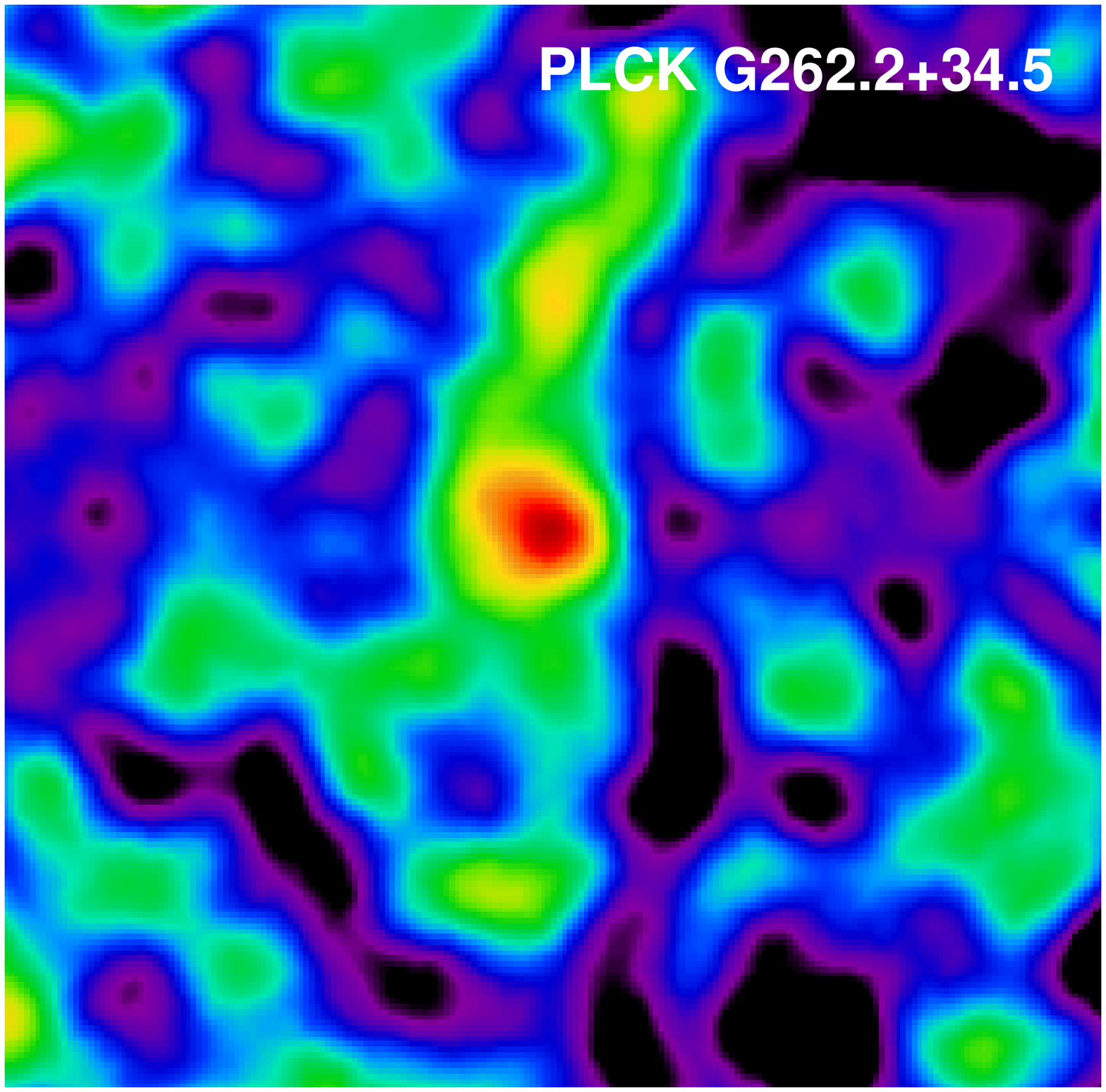}%
\hspace{5mm}
\includegraphics[scale=1.,angle=0,keepaspectratio]{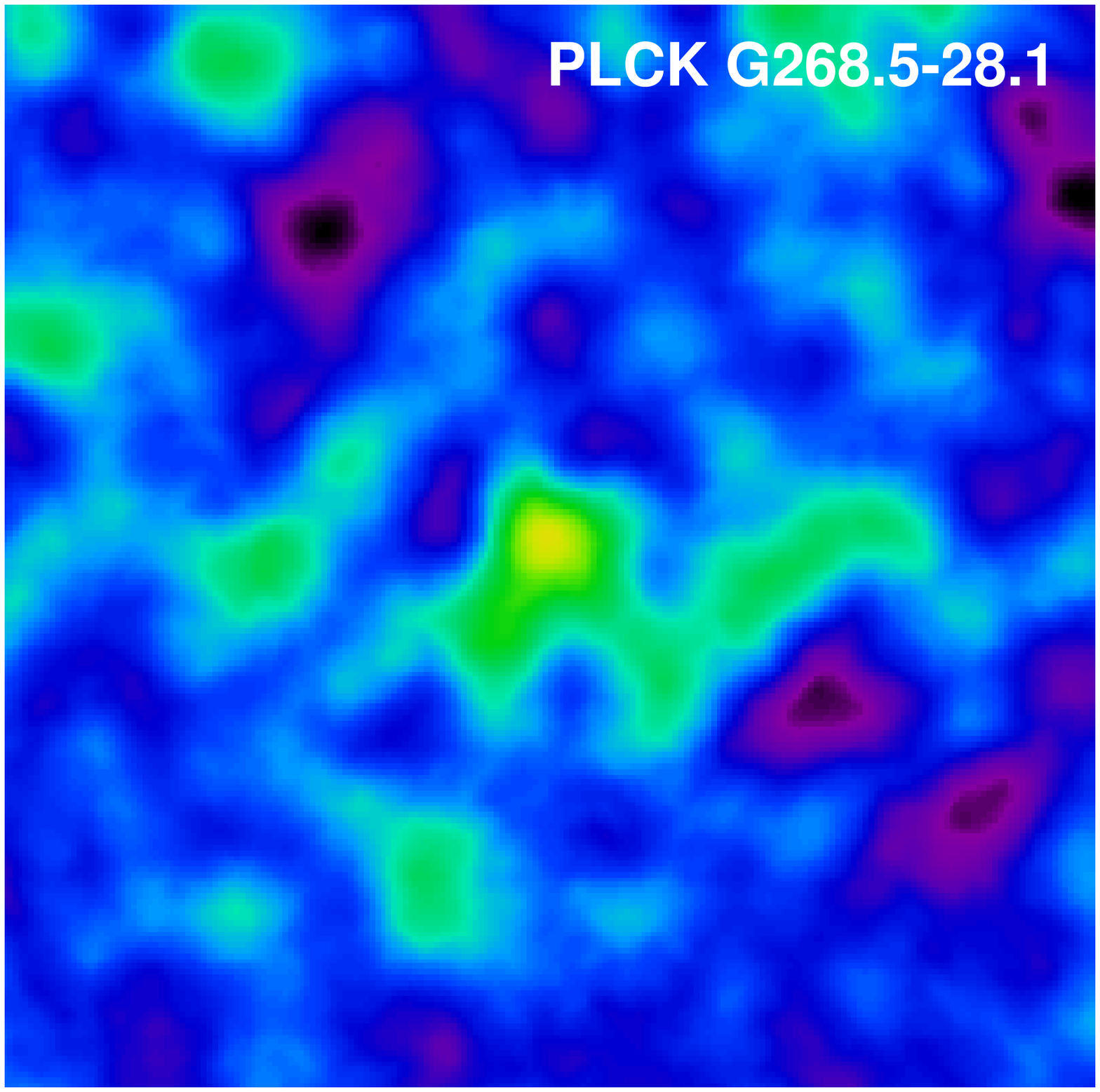}
} 
\end{minipage}
\begin{minipage}[t]{0.85\textwidth}
\vspace{1mm}
\resizebox{\hsize}{!} {
\includegraphics[scale=1.,angle=0,keepaspectratio]{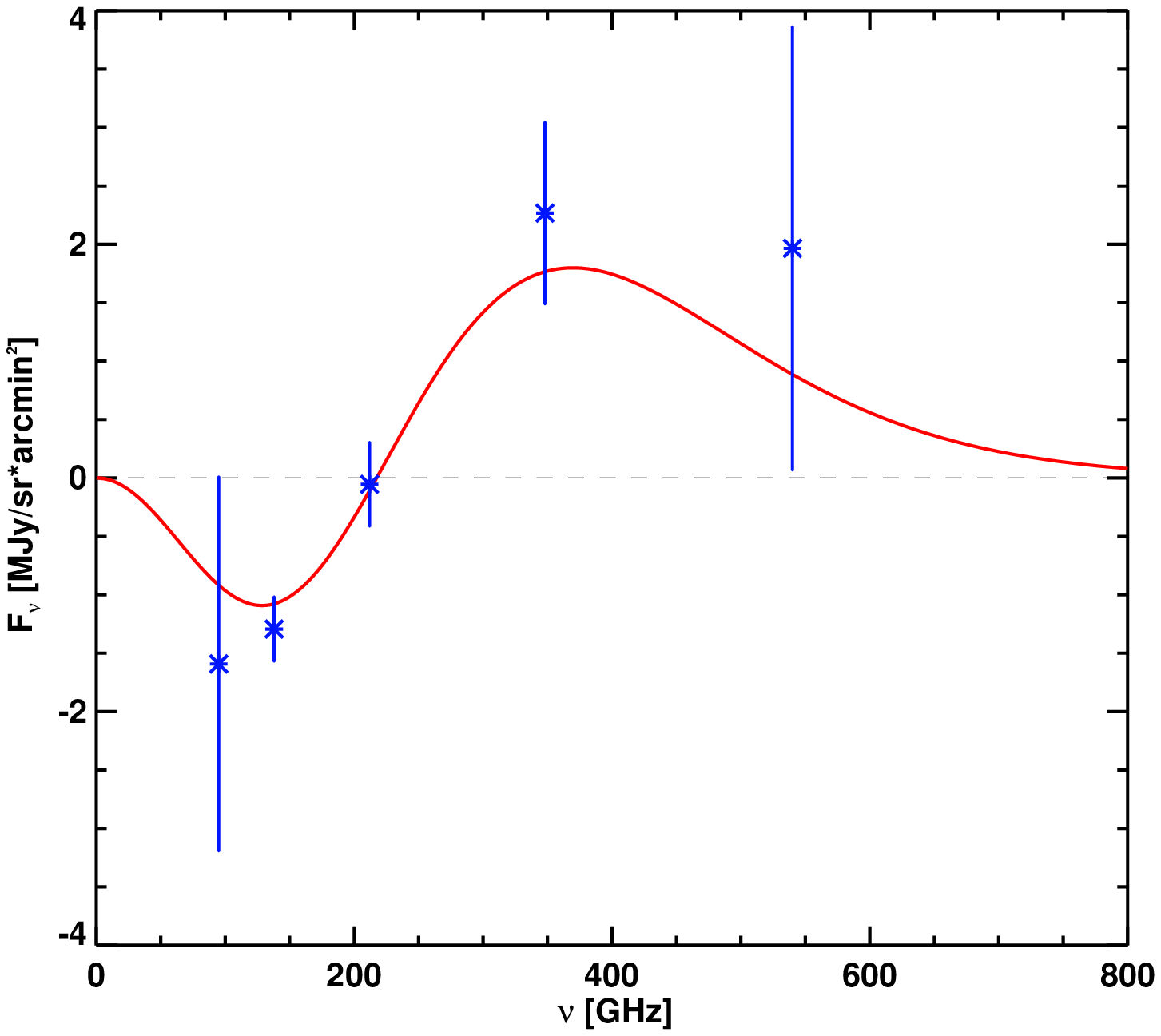}%
\hspace{5mm}
\includegraphics[scale=1.,angle=0,keepaspectratio]{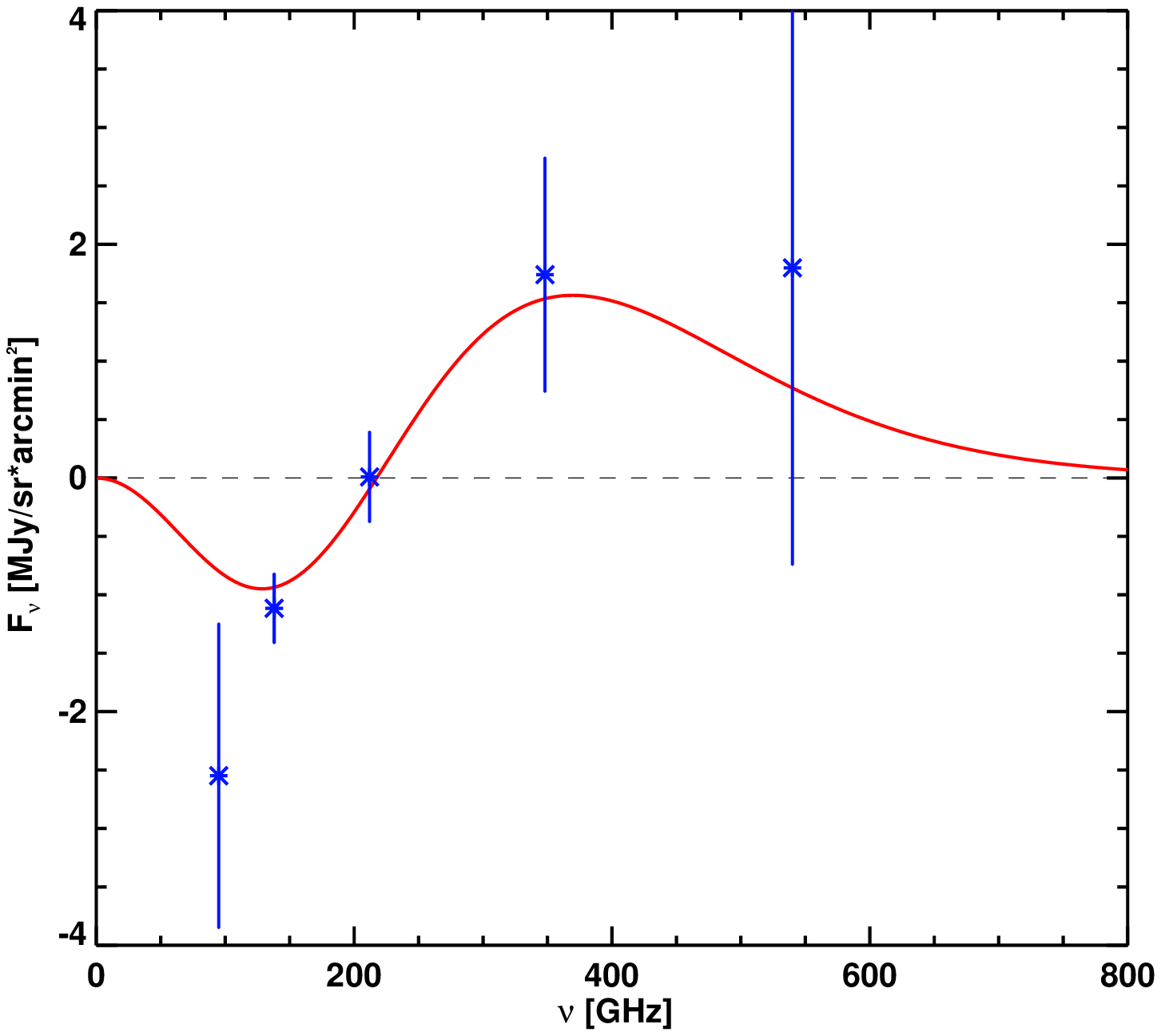}%
\hspace{5mm}
\includegraphics[scale=1.,angle=0,keepaspectratio]{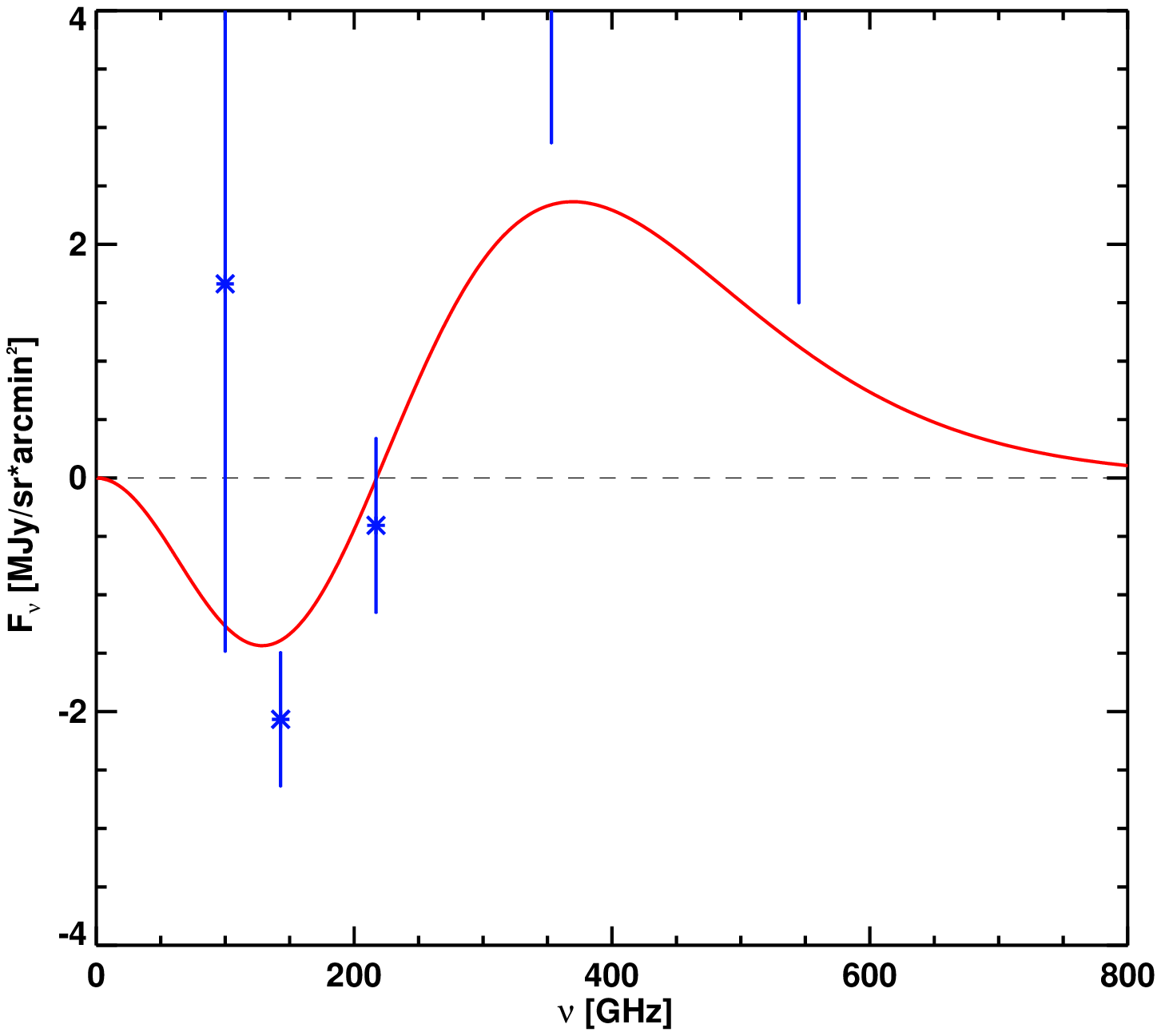}
}
\end{minipage}
\caption{{\footnotesize Illustration of the three SZ quality grades as defined in Sec.~\ref{sec:sel}. From left to right the three quality cases are: $Q_{SZ}=A,B,C$. The top row shows a $100 \arcmin\times100 \arcmin$ SZ map with a spatial resolution of $10\arcmin$,  centred on the candidate position, derived using the MILCA reconstruction method \citep{hur10}. The colour table is identical for all clusters, with the Compton $y$ parameter spanning the range $[-3\times10^{-6},1\times10^{-5}]$. The bottom row shows the associated SZ spectrum from aperture photometry measurements within $R_{500}$ (see text for details). The red line is the SZ spectrum normalised to the $\YSZ$ value obtained from MMF blind detection. \label{fig:qz}}}
\end{figure*}
%________________________________________________________________

\section{ Sample selection}
\label{sec:sel}

The present candidates were chosen from the catalogue of detections in the all-sky maps from the first ten months of the \planck\ survey. This same catalogue  was  used  for the construction of the ESZ sample, for which the reference method for blind cluster searches was the matched multi-frequency filter 
`MMF3', developed by \citet[][]{mel06}. Complementary searches were also performed with an independent implementation of the MMF method and with the PowellSnakes algorithm \citep[PWS;][]{car09b,car11}. As described in \citet{planck2011-5.1a}, candidates then underwent a validation process, including internal SZ quality checks. The first part of this process included an initial quantitative assessment of the blind SZ signal detection, based on the $\mathrm{S/N}$ ratio and the number of methods blindly detecting the candidate, $N_{\rm det}$. 

The quality of the SZ signal cannot simply be reduced to a single global ${\rm S/N}$ value. It depends not only on the intrinsic cluster SZ signal, but also on the detailed local properties of the various noise components, i.e, the background (e.g., CIB, CMB) and foreground environments (e.g., galactic dust, synchrotron, free-free emissions). Therefore, beyond the quantitative criteria stated above, we also performed a {\it qualitative} assessment of the SZ signal based on visual inspection of SZ maps and spectra.

We first examined frequency maps, using both raw maps made directly from the \planck\ all sky data, and maps that had been cleaned of dust emission. We used IRIS-100$\mu$m \citep{miv05} and \planck\ HFI-857~GHz maps as dust templates, and the `dust-cleaned' HFI-217~GHz map as a CMB template. These frequency maps were investigated for strong foreground dust contamination and the presence of submillimetre sources on the high frequency side. Radio source contamination and CMB residuals were searched for at low frequencies. In addition to the frequency maps, reconstructed SZ maps were built using three different reconstruction methods based on Independent Linear Component (ILC) analysis \citep[e.g.,][]{hur10}. Finally, SZ spectra were built from the SZ flux estimation at each \planck\ frequency. Spectra were estimated both from the best detection outputs and also directly from aperture photometry on CMB- and dust-cleaned maps.

On the basis of the frequency maps, the reconstructed SZ maps, and the spectra for each cluster, we then defined three SZ quality grades, $Q_{\rm SZ}$:
\begin{itemize}
\item{$Q_{\rm SZ}=A,$} if all the following criteria are fulfilled:
\begin{itemize}
\item  Clear compact SZ source detected in the SZ map. 
\item Obvious measurements of the SZ decrement at least at 143\,GHz or 100\,GHz.
\item Low dust contamination (i.e., no increase in the 353~GHz and 545~GHz fluxes in the SZ spectrum or residual dust emission or submillimetre point sources in the frequency map), and a reasonable detection at 353~GHz. 
\item No radio source contamination (checked in LFI maps) or CMB confusion (checked in the HFI 217\,GHz map). 
\end{itemize}
\item{$Q_{\rm SZ}=B,$} if all the following criteria are fulfilled:
\begin{itemize}
\item Visible SZ detection in the SZ map or significant measured SZ signal at 143\,GHz. The 100~GHz signal can be more noisy.
\item Dust emission well subtracted but for the effect of point source contamination at the cluster location, (i.e., increase of the 353~GHz and possibly the 545~GHz fluxes in the SZ spectrum or residual dust emission or submillimetre point sources in the frequency map) resulting in large uncertainties for dust emission removal.
\item No radio source contamination or CMB confusion. 
\end{itemize}
\item{$Q_{\rm SZ}=C,$} if any of the three following criteria are fulfilled:
\begin{itemize}
\item Weak SZ spectral signature (due to large error bars or to inconsistent spectral shape) or visible signal in noisy SZ maps.
\item Strong dust contamination (i.e., high 353~GHz and 545~GHz fluxes in the SZ spectrum or residual dust emission or submillimetre point sources in the frequency map). 
\item Possible contamination by radio sources seen down to the LFI-70\,GHz channel.
\end{itemize}
\end{itemize}
The three cases are illustrated in Fig.~\ref{fig:qz}. 
These criteria were checked using the maps and spectra obtained with the different methods described above. Convergence between methods helped us to define the quality grade for each candidate.

%________________________________________________________________
\begin{table*}[t]
\begingroup
\caption{ {\footnotesize Observation log of the \xmm\ validation follow-up.  Column (1):  \planck\ source name.  Column (2)--(3): Signal-to-noise ratio of the \planck\ cluster candidate detection with the MMF3 algorithm in the v4.1 \planck-HFI maps, number of methods blindly detecting the candidate. Column (4): quality grade of the SZ detection (A is best). Columns (5) and (6): Right ascension and declination  of the \planck\ source (J2000). Columns (7)--(10):  \xmm\ observation identification number, filter used, on-source exposure time with the \pn\ camera and fraction of useful time after cleaning for periods of high background due to soft proton flares (\mos\ and \pn\ camera, respectively). Asterisked objects denote observations affected by a high background level. Column (11): Confirmed clusters are flagged. }\label{tab:obs}}
\centering  
\begin{tabular}{lrrrrrrcccc}
\toprule
\multicolumn{1}{c}{Name} & \multicolumn{1}{c}{$\textrm{S/N}$} & \multicolumn{1}{c}{$\textrm{$N_{det}$}$} & \multicolumn{1}{c}{$\textrm{$Q_{SZ}$}$} &  \multicolumn{1}{c}{RA$_{\rm SZ}$} & 
\multicolumn{1}{c}{DEC$_{\rm SZ}$}  & \multicolumn{1}{c}{OBSID} & 
\multicolumn{1}{c}{filter} & \multicolumn{1}{c}{$t_{\rm exp}$} &
\multicolumn{1}{c}{Clean fraction} & \multicolumn{1}{c}{Confirmed}\\
\noalign{\smallskip}
\multicolumn{1}{c}{} & \multicolumn{1}{c}{} &  \multicolumn{1}{c}{}&  \multicolumn{1}{c}{} & \multicolumn{1}{c}{(deg)} & 
\multicolumn{1}{c}{(deg)}  & \multicolumn{1}{c}{}  &  
\multicolumn{1}{c}{} & \multicolumn{1}{c}{(ks EPN)} &
\multicolumn{1}{c}{(MOS/EPN)} & \multicolumn{1}{c}{} \\
\midrule
PLCK~G060.1+15.6& 5.3& 3		&	 B	&280.279&  30.930& 658200901&MMM&10.0&0.4/0.3&    Y\\
PLCK~G200.9$-$28.2& 5.2&  2 	&A	&72.564&  $-$3.002& 658200801&TTT&11.2&0.8/0.1*&    Y\\
PLCK~G235.6+23.3& 5.2&  3		&B	&134.032&  $-$7.719& 658201301&TTT&10.7&0.4/0.2&    Y\\
PLCK~G113.1$-$74.4& 5.1&  2	&C	&10.161& $-$11.706& 658200601&TTT&10.0&1.0/0.7& $\ldots$ \\
PLCK~G262.2+34.5& 5.1& 3 		&B	&158.596& $-$17.342& 658201001&MMT&12.4&0.9/0.5&    Y\\
PLCK~G268.5$-$28.1& 5.1&  3	&C	&92.855& $-$59.611& 658201101&TTT&11.0&0.8/0.6*&    Y\\
PLCK~G266.6$-$27.3& 5.0&  3	&B	&94.027& $-$57.791& 658200101&TTT&10.0&0.8/0.2*&    Y\\
PLCK~G019.1+31.2& 5.0& 3		&A	&249.143&   3.153& 658200301&TTT&10.0&1.0/0.8&    Y\\
PLCK~G193.3$-$46.1& 4.9&  3	&B	&53.960&  $-$6.985& 658200401&TTT&12.1&0.8/0.6&    Y\\
PLCK~G234.2$-$20.5& 4.7&  3	&C	&92.747& $-$27.544& 658201201&TTT&13.7&1.0/0.8&    Y\\
PLCK~G210.6+17.1& 4.6& 2		&C	&117.214&   9.688& 658200501&TTT&11.7&1.0/0.8&    Y\\
\bottomrule
\end{tabular}
\endPlancktable 
 \endgroup
\end{table*}
%_____________________________________________________________

We chose candidates to examine our internal SZ quality assessment by exploring lower quality detections than in our previous publications. On the basis of a candidate list detected by at least two algorithms\footnote{Note that the same candidate list was used to define the ESZ.}, we selected eleven candidates detected  at  $4.5\!<\!\textrm{S/N}\!<\!5.3$ with the MMF3 algorithm. Here we are sampling a lower $\textrm{S/N}$ regime than in our previous validation run (for which $5.1\!<\!{\rm S/N}\!<\!10.6$) or in the ESZ sample (for which ${\mathrm S/N}>6$). 
To investigate the pertinence of our SZ quality grade definitions, we selected  typical cluster candidates from the three categories, in the following proportions: two, five, and four, respectively, for  $Q_{SZ}=$~A, B and C. The SZ properties of the candidates are given in Table~\ref{tab:obs}.  Note that the objects in no way constitute a complete or even statistically representative sample. Hence, we cannot use them to draw any quantitative conclusions regarding, for example, the purity of the parent catalogue. 

Two of the three lowest ${\rm S/N}$ candidates, PLCK~G193.3$-$46.1 and  PLCK~G210.6+17.1 fall in the Sloan Digital Sky Survey (SDSS) area.  They have no counterpart in published SDSS cluster catalogues, but our dedicated algorithm search for galaxy over-densities \citep{fro11} indicated that they were each possibly associated with a $z>0.5$ cluster. Inclusion of these two targets allowed us to further test SDSS-based confirmation at high $z$.

%________________________________________________________________
\begin{figure*}[tbp]
\center
\includegraphics[scale=1.,angle=0,keepaspectratio,height=0.77\textheight, clip]{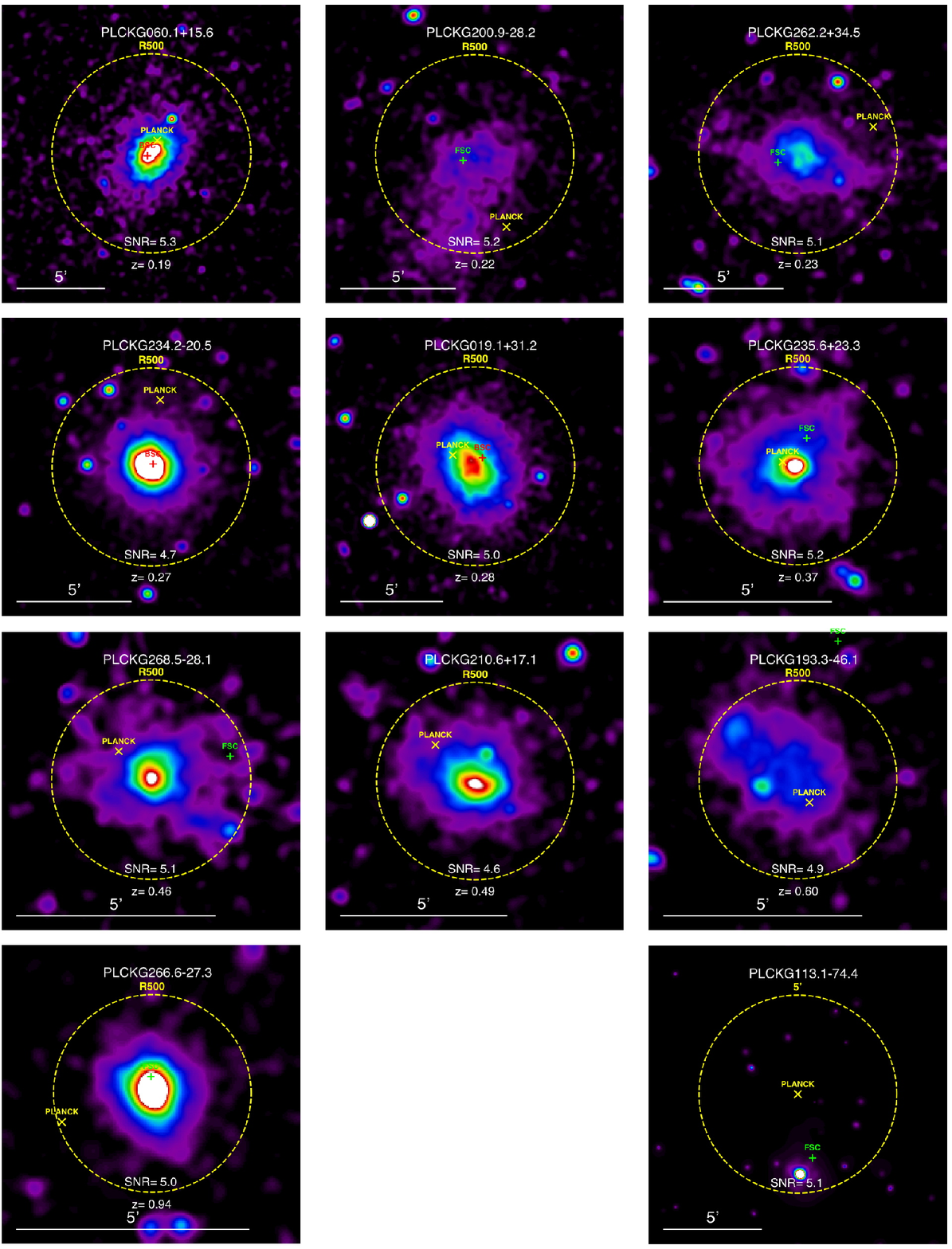}
\caption{{\footnotesize \xmm\  $[0.3$--$2]\,\keV$ energy band cluster candidate images. North is up and east to the left. The bottom right-hand panel shows the image of the false \planck\ candidate.  For confirmed clusters, image sizes are $3 \theta_{500}$ on a side, where $\theta_{500}$ is estimated from the $\Mv$--$\YX$ relation (Eq.~\ref{eq:yx}). Images are corrected for surface brightness dimming with $z$, divided by the emissivity in the energy band, taking into account galactic absorption and instrument response, and scaled according to the self-similar model. The colour table is the same for all clusters, so that the images would be identical if clusters obeyed strict self-similarity. The majority of the objects show evidence for significant morphological disturbance.  A yellow cross indicates the \planck\ position and a red/green plus sign the position of a \rass-BSC/FSC source, respectively.}}
 \label{fig:gal}
\end{figure*}
%________________________________________________________________

%________________________________________________________________
\begin{table*}[t]
\caption{{\footnotesize X-ray and SZ properties of the confirmed \planck\ sources. 
 Columns (2)--(3): Right ascension and declination of the peak of the X--ray emission (J2000). Column (4): redshift from X-ray spectral fitting.  Column (5): Quality flag for the X-ray redshift measurement (2 is best). Column (6): Total \epic\  count rates  in the $[0.3$--$2]\,\keV$ band,  within the maximum radius of detection given in column (7).  Columns (8)--(14): $\Rv$ is the radius corresponding to a density contrast of 500,  estimated iteratively from the \MYX\ relation  (Eq.~\ref{eq:yx}), where $Y_{\rm X}=\Mgv\TX$ is the product of the gas mass within $\Rv$ and the spectroscopic temperature $\TX$, and $\Mv$ is the total mass within $\Rv$.  $\LX$ is the luminosity within $\Rv$ in the $[0.1-2.4]$ keV band.  $\Yv$ is the spherically integrated Compton parameter measured with \planck, centred on the X--ray peak, interior to the $\Rv$ estimated with the X-ray observations.
  }
\label{tab:xray}} 
\resizebox{\textwidth}{!} {
\begin{tabular}{lrrllrrrrrrrrc}
\toprule
\multicolumn{1}{c}{Name} & 
\multicolumn{1}{c}{RA$_{\rm X}$} & \multicolumn{1}{c}{DEC$_{\rm X}$} & 
\multicolumn{1}{c}{$z_{\rm Fe}$} & \multicolumn{1}{c}{$Q_{\rm z}$} &
\multicolumn{1}{c}{$R$} & \multicolumn{1}{c}{$\theta_{\rm det}$} &
\multicolumn{1}{c}{$R_{500}$} & \multicolumn{1}{c}{$\TX$} &
\multicolumn{1}{c}{$M_{\rm gas,500}$} & \multicolumn{1}{c}{$D_{\rm A}^{-2}\,C_{\rm XSZ}\,\YX$} &
\multicolumn{1}{c}{$Y_{500}$} & \multicolumn{1}{c}{$M_{500}$} &
\multicolumn{1}{c}{$\LX$} \\
\noalign{\smallskip}
\multicolumn{1}{c}{} &
\multicolumn{1}{c}{[h:m:s]} &\multicolumn{1}{c}{[d:m:s]} & 
\multicolumn{1}{c}{} &
\multicolumn{1}{c}{} & \multicolumn{1}{c}{$[{\rm cts\,s^{-1}}]$} & \multicolumn{1}{c}{$[\arcmin]$} &
\multicolumn{1}{c}{[Mpc]} & \multicolumn{1}{c}{[keV]} &
\multicolumn{1}{c}{$[10^{14}\,{\rm M_{\odot}}]$} & \multicolumn{1}{c}{$[10^{-4}\,{\rm arcmin^2}]$} &
\multicolumn{1}{c}{$[10^{-4}\,{\rm arcmin^2}]$} & \multicolumn{1}{c}{$[10^{14}\,{\rm M_{\odot}}]$} &
\multicolumn{1}{c}{$[10^{44}\,{\rm erg\,s^{-1}}]$} \\
\midrule
PLCK~G060.1+15.6		&{18:41: 08.5}&{+30:55:03.4}	&$0.19$&2&$ 1.52\pm0.03$&  5.4&$1.07\pm0.02$&$ 6.9\pm0.6$&$ 0.40\pm0.02$				&$10.9\pm1.4$&$ 9.0\pm2.2$&$ 4.3\pm0.3$&$ 2.65\pm0.06$\\
 PLCK~G200.9$-$28.2	&{04:50:20.9}&{$-$02:56:57.6}	&$0.22$&2&$ 0.38\pm0.03$&  3.1&  $0.91\pm0.04$&$ 4.5\pm0.7$&$ 0.27\pm0.02$			&$3.8\pm0.8$&$ 4.3\pm3.4$&$ 2.7\pm0.3$&$ 0.99\pm0.04$\\
 PLCK~G235.6+23.3	&{08:56:05.9}&{$-$07:43:15.3}	&$0.37$&2&$ 0.67\pm0.02$&  4.9& $ 0.95\pm0.02$&$ 4.2\pm0.3$&$ 0.51\pm0.01$			&$3.2\pm0.3$&$ 6.7\pm1.5$&$ 3.6\pm0.2$&$ 3.50\pm0.09$\\
 PLCK~G262.2+34.5	&{10:34:36.2}&{$-$17:21:40.2}	&$0.21$&1$^\dagger$&$ 0.83\pm0.02$&  6.0& $ 0.91\pm0.02$&$ 4.0\pm0.3$&$ 0.30\pm0.01$	&$3.8\pm0.4$&$ 6.5\pm2.0$&$ 2.6\pm0.1$&$ 1.42\pm0.03$\\
 PLCK~G268.5$-$28.1	&{06:11:18.9}&{$-$59:37:23.2}	&$0.47$&1$^\ast$&$ 0.37\pm0.01$&  3.3& $ 0.83\pm0.03$&$ 3.5\pm0.4$&$ 0.39\pm0.02$		&$1.6\pm0.3$&$ 3.9\pm0.9$&$ 2.7\pm0.3$&$ 3.39\pm0.14$\\
 PLCK~G266.6$-$27.3	&{06:15:52.1}&{$-$57:46:51.6}	&$0.94$&2&$ 0.52\pm0.02$&  2.3& $ 0.98\pm0.03$&$10.5\pm1.5$&$ 1.04\pm0.02$			&$7.0\pm1.0$&$ 4.3\pm0.9$&$ 7.8\pm0.6$&$22.7\pm0.8$\\
 PLCK~G019.1+31.2	&{16:36:29.8}&{+03:08:37.5}	&$0.28$&2&$ 1.88\pm0.02$&  4.1&$ 1.24\pm0.01$&$ 7.7\pm0.4$&$ 0.95\pm0.03$			&$16.0\pm1.3$&$15.0\pm2.3$&$ 7.2\pm0.3$&$ 6.24\pm0.09$\\
 PLCK~G193.3$-$46.1	&{03:35:51.4}&{$-$06:58:32.8}	&$0.59$&1$^\ddagger$&$ 0.32\pm0.01$&  2.9& $ 0.99\pm0.03$&$ 6.0\pm0.7$&$ 0.79\pm0.06$	&$4.3\pm0.8$&$ 7.1\pm1.4$&$ 5.3\pm0.5$&$ 5.04\pm0.17$\\
 PLCK~G234.2$-$20.5	&{06:11:01.2}&{$-$27:35:31.7}	&$0.27$&2&$ 1.71\pm0.02$&  4.8&$ 1.07\pm0.01$&$ 6.1\pm0.3$&$ 0.53\pm0.01$			&$7.4\pm0.4$&$ 5.2\pm1.8$&$ 4.5\pm0.1$&$ 4.87\pm0.05$\\
 PLCK~G210.6+17.1	&{07:48:46.6}&{+09:40:10.6}	&$0.48$&2&$ 0.67\pm0.01$&  4.2&$ 1.07\pm0.02$&$ 7.0\pm0.4$&$ 0.79\pm0.02$			&$6.0\pm0.5$&$ 5.5\pm1.6$&$ 5.8\pm0.3$&$ 6.21\pm0.11$\\
\bottomrule
\end{tabular} }
\tablefoot{
$^\dagger$ Other possible $z_{\rm Fe}$:   $0.02, 0.82$; The best estimate, $z_{\rm Fe}=0.21$ is consistent with the optical photometric redshift.   \\
$^\ddagger$ Other possible $z_{\rm Fe}$: $ 0.19, 0.82$; The best estimate, $z_{\rm Fe}=0.59$ is consistent with the optical photometric redshift.  \\
$^\ast$ Other possible $z_{\rm Fe}$:  $0.12, 0.87, 1.20$ }
\\
\normalsize
 \end{table*}
%________________________________________________________________

\section{\xmm\ observations}

The data analysis and validation procedure is described extensively by  \citet{planck2011-5.1b}. We present only a brief summary in this section.

\subsection{Observations and data reduction}
The candidates were observed between December 22, 2010 and May 16, 2011. The observation identification number and observation setup are summarised in Table~\ref{tab:obs}. The nominal setup used the THIN filters (unless optical loading needed to be avoided) and extended full frame (EFF)  mode for the \pn\  camera. 

Calibrated event lists were produced with v11.0 of the \xmm\ Science Analysis System.  Data that were affected by periods of high background due to soft proton flares were omitted from the analysis; clean observing time after flare removal is given  Table~\ref{tab:obs}. Three observations are affected by high background levels: PLCK~G268.5$-$28.1, PLCK~G200.9$-$28.2 and PLCK~G266.6$-$27.3. The data treatment for the latter cluster is fully described in  \citet{planck2011-5.1c}.  For PLCK~G268.5$-$28.1 and PLCK~G200.9$-$28.2,  the particle background after flare cleaning is $2$ and $1.7$ times higher than nominal for the \pn\ camera, respectively. The \pn\ data were thus discarded for the spectroscopic analysis, as this is very sensitive to the background estimate. 

The cleaned data were {\sc pattern}-selected and corrected for vignetting as described in \citet{pra07}.  Bright point sources were excised from the data. The  background treatment is as described in \citet{pra10}. 
In the spectroscopic analysis, the cluster component was modeled with an absorbed thermal emission model (\mekal) with a hydrogen column density fixed at the 21-cm value of \citet{dic90}. 

\subsection{Candidate confirmation}

The confirmation status of each \xmm\ observation is given in Table~\ref{tab:obs} and the \xmm\ images  are shown in Fig.~\ref{fig:gal}.  Of eleven targets, ten candidates are  {\it bona fide} clusters.  In each case, the  extended nature of the X-ray source, clearly detected within the \planck\ position error box, was  confirmed by comparing the surface brightness profile with the \xmm\ point spread function (PSF).  The consistency between the SZ and X--ray properties (Sec.~\ref{sec:outxsz}) provided the final confirmation check.  The total \epic\  count rates  in the $[0.3$--$2]\,\keV$ band of each cluster and the maximum radius of detection are given in  Table~\ref{tab:xray}.

The offset between the  X--ray position and the \planck\ position (Fig.~\ref{fig:gal})  is similar to that observed for known clusters in the ESZ sample \citep{planck2011-5.1a} or for candidates that have previously been confirmed with \xmm\ \citep{planck2011-5.1b}.
The median offset is $1\farcm6$, characteristic of the \planck\ reconstruction uncertainty, which peaks around $2\arcmin$ \citep{planck2011-5.1a,planck2011-5.1b} and is driven by the spatial resolution of the instruments. The largest offset is  $3\farcm4$ or $0.8\Rv$. This offset  is observed for PLCK G200.9$-$28.2, a highly disturbed cluster with a flat X-ray  morphology (Fig.~\ref{fig:gal}), for which a true physical offset between the X-ray and SZ signal may also contribute.

One candidate, PLCK~G113.1$-$74.4, proved to be a false detection (Fig.~\ref{fig:gal}, last panel) as no extended source is detected within the \planck\ position error. The surface brightness profile of the \rass\ Faint Source Catalogue source, detected about $5\arcmin$ South of the \planck\ position, is consistent with that of a point source. 

\subsection{ Redshift and physical parameter estimates}

To estimate the redshift from the X-ray data, $z_{\rm Fe}$, we extracted a spectrum within a circular region corresponding to the maximum significance of the X--ray detection.  Since the centroid of the Fe--K  line complex  depends on the temperature, the redshift was determined from a thermal model fit to the full spectrum in the $[0.3$--$10]\,\keV$ band,  as described in detail in \citet{planck2011-5.1b}. The quality of the redshift estimate was characterised by the quality flag  $Q_{\rm z}$ as defined in \citet{planck2011-5.1b}. The redshift of most clusters is well constrained ($Q_{\rm z}\!\!=\!\!2$). Three clusters,PLCK~G193.3$-$46.1, PLCK-G262.2+34.5 and PLCK~G268.5-28.1,  have ambiguous $z_{\rm Fe}$ estimates ($Q_{\rm z}\!\!=\!\!1$). They exhibit several $\chi^2$ minima in the $\kT$--$z_{\rm Fe}$ plane that do not differ at the $68\%$ confidence level (see Sect.~\ref{sec:zxsz} for further discussion).  For these systems we used the redshift corresponding to the most significant $\chi^2$ minimum, listed in Table~\ref{tab:xray}.  For PLCK~G193.3$-$46.1 and PLCK~G262.2+34.5,  this redshift corresponds to the  optical photometric redshift subsequently derived  from SDSS data (Sect.~\ref{sec:outqual}) and our optical follow-up (Sec.~\ref{sec:eso}), respectively. The uncertainty on the redshift  is not propagated through the physical parameter estimation procedure discussed below.  The statistical uncertainty on $z_{\rm Fe}$ is small for the $Q_{\rm z}\!\!=\!\!2$ systems. The physical parameters for  $Q_{z}\!\!=\!\!1$ systems, especially PLCK~G268.5-28.1, are less robust and should be treated with caution.

 We then derived the  gas density profile  of each cluster from the surface brightness profile, using the regularised deprojection and PSF--deconvolution technique developed by \citet{cro06}.
Global cluster parameters were estimated self-consistently within $\Rv$ via iteration about the $\Mv$--$\YX$ relation of \citet{arn10} assuming standard evolution:
\begin{equation}
  E(z)^{2/5}\Mv = 10^{14.567 \pm 0.010} \left[\frac{\YX}{2\times10^{14}\,{\msol}\,\keV}\right]^{0.561 \pm 0.018}\,\msol.
  \label{eq:yx}
  \end{equation}
The quantity $\YX$, introduced by \citet{kra06},  is defined as the product  of  $\Mgv$, the gas mass within $\Rv$, and $\TX$, where the latter is the spectroscopic temperature measured in the $[0.15$--$0.75]\,\Rv$ aperture. In addition, $L_{500}$, the X-ray luminosity inside $\Rv$, was calculated as described in \citet{pra09}. The errors on $\Mv$ given in the table correspond to statistical uncertainties only. Additional errors due to scatter around the relation (around $7\%$ from simulations) and uncertainties on the relation itself are not taken into account.

The SZ flux was then re-extracted, calculating $\Yv$ with the X-ray position and size $\Rv$ fixed to the refined values derived from the high-quality \xmm\ observation. The X-ray properties of the clusters and resulting refined $Y_{500}$ values are listed in Table~\ref{tab:xray}. For most cases, the blind values  are consistent  with the recomputed  $\Yv$, within the errors. However, as found in our previous studies \citep{planck2011-5.1a,planck2011-5.1b}, there is a trend of SZ flux overestimation with size overestimation. For the present sample, the blind values are overestimated by  a median factor of 1.3 for the size and 1.4 for $\Yv$.

We have checked for possible AGN contamination using the NVSS  \citep[at 1.4\,GHz][]{con98} and SUMSS \citep[at 0.84\,GHz][]{boc99} catalogues. A relatively bright radio source ($560\,{\rm mJy}$) is found in the vicinity of  PLCK G193.3$-$46.1 (at $7\farcm6$ offset). However, LFI data do not show any significant signal so the source must have a steep spectrum.  No other radio sources are found in any other candidates. We conclude that no  significant contamination of the SZ signal is expected in any of the clusters.  However, we cannot exclude the presence of radio faint AGN within each cluster area. Although they could contaminate the X-ray signal if present, the brightest X--ray sources are resolved and excised from the X--ray analysis. 
 
 %________________________________________________________________
\begin{figure}[t]
\center
\includegraphics[width=0.85\columnwidth]{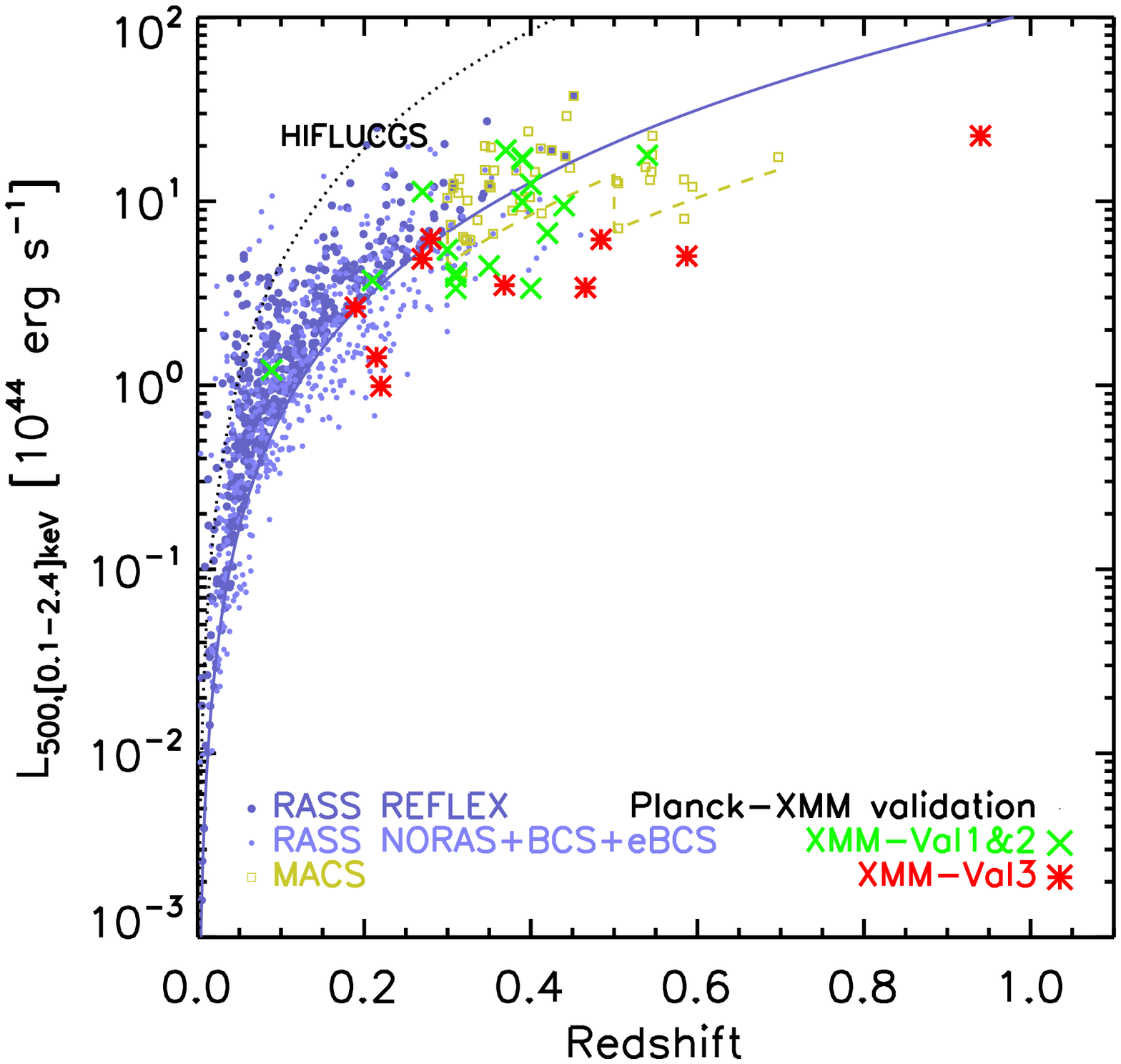}
\caption{ {\footnotesize 
The new SZ-discovered \planck\ objects (red and green symbols) compared to clusters from the \rosat\  All-Sky Survey catalogues in the \Lxz\ plane.  The X--ray luminosity  is calculated  in the $[0.1$--$2.4]\,\keV$ band. Catalogues shown are \reflex\ \citep{boe04}, \noras\ \citep{boe00}, \bcs\ \citep{ebe98}, \ebcs\ \citep{ebe00} and \macs\  \citep{ebe07}. The solid line is the REFLEX flux limit, the dotted line is the HIFLUCGS flux limit of $2 \times 10^{-11}\ergscm$ and the dashed line is from  the \macs\ flux limits. 
} \label{fig:lxz}} 
\end{figure}
%________________________________________________________________

\section{\xmm\ validation outcome}

\subsection{ \planck\ sensitivity}

The present validation run clearly demonstrates the capability of \planck\ to detect clusters of a wide range of masses up to high $z$.  All targets in this run  fall below the \rass\ X-ray flux limit. This is illustrated in  Fig.~\ref{fig:lxz}, where the new clusters are shown in the \Lxz\  plane. They are plotted together with the clusters from large catalogues based on \rass\  data outside the Galactic Plane and the clusters confirmed in previous \xmm\ validation observations (hereafter {\it XMM}-Val1$\&$2 ). The new sample covers a wide range of redshift, $0.2<z<1$. It includes two clusters at $z>0.5$ and the first cluster blindly detected by \planck\ at $z\sim1$ \citep[see][for a detailed discussion of this cluster]{planck2011-5.1c}. 
The new clusters are less X--ray bright, at a given $z$, than those previously confirmed with \xmm. This is not surprising, since we are probing a lower ${\rm S/N}$, thus less massive, cluster candidate regime. The new clusters all lie  below the \rass\ survey flux limits, even that of the most sensitive survey (\macs). The mass estimates range from as low as $\Mv=2.7\pm0.2\times10^{14}\msol$ for the nearby $z=0.22$ cluster,  PLCK~G200.9$-$28.2, to $\Mv=7.8\pm0.6\times10^{14}\msol$ for  PLCK~G266.6$-$27.3 at $z=0.97$.   Interestingly, the two clusters are detected at very similar ${\rm S/N}$. This reflects the \planck\ selection function, which depends on the integrated SZ flux, i.e., on the size and redshift of the cluster. \planck\  can detect both (1) low $z$, low mass clusters with large angular extent, and (2) compact high $z$, high mass objects. Consequently, the mass detection threshold of the \planck\ survey increases with redshift (at least in the redshift range probed by the present sample).
 
%________________________________________________________________
%% Figure: gas density profiles
%%
\begin{figure}[t]
\center
\includegraphics[width=0.85\columnwidth]{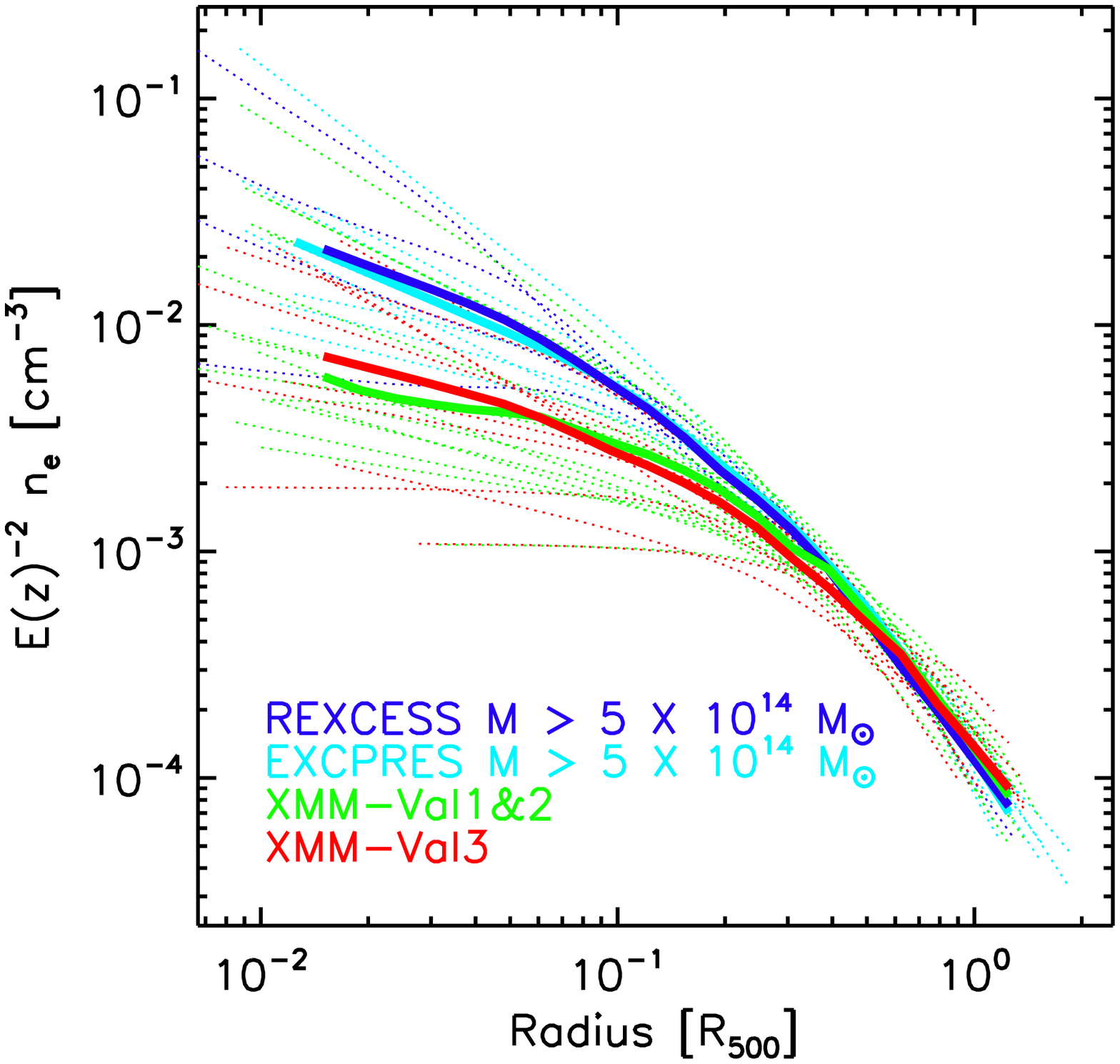}
\caption{{\footnotesize 
Regularised  scaled density profiles of the new confirmed \planck\ SZ clusters with redshift estimates ($0.2<z<0.97$, red lines). They are compared to those of similar mass systems from the representative X-ray samples \rexcess\ \citep[][blue lines]{boe07}, \excpres\ (Arnaud et al., in prep., cyan lines), and the new clusters at lower redshift ($0.09<z<0.54$) and higher SZ flux confirmed in previous validation runs \citep[][green lines]{planck2011-5.1b}. The thick lines denote the mean scaled profile for each sub-sample. }}\label{fig:ne}
\end{figure}
%________________________________________________________________

\subsection{Candidate quality assessment}
\label{sec:outqual}

Only one of the eleven  candidates, PLCK~G113.1$-$74.4, is false.  It is noteworthy that its ${\rm S/N}$ is the fourth-highest of the sample (${\rm S/N}=5.1$). While this is rather high, its actual SZ detection falls into the lowest quality category, $Q_{\rm SZ}=C$, an indication of the importance of the quality grades defined in Sect.~\ref{sec:sel}, in addition to the ${\rm S/N}$ ratio. The other three $Q_{\rm SZ}=C$ candidates are confirmed, including  PLCK~G268.5$-$28.1, detected at the same ${\rm S/N}$ as the false candidate, and PLCK~G234.2$-$20.5, detected at ${\rm S/N}=4.7$, the second lowest ${\rm S/N}$ of the sample. Both clusters are detected by all three SZ detection methods, whereas the false candidate is only detected with the two MMF methods and not with the PWS algorithm.  As expected, the probability that a candidate is a true cluster increases with $N_{\rm det}$. 

As previously noted by \citet{planck2011-5.1b},  association of a cluster candidate with a \rass\ source  within the \planck\ position uncertainty is not, by itself, sufficient for confirmation. The false candidate, PLCK~G113.1$-$74.4, was associated with a \rass/FSC source that eventually proved to be a point source.  

The lowest ${\rm S/N}$ candidate of all, PLCK~G210.6+17.1, is confirmed, whereas it was detected by only two SZ detection  methods and lies in the lowest quality category,  $Q_{\rm SZ}=C$. However, it is one of the two clusters that was flagged by our SDSS detection algorithm as being possibly associated with an SDSS cluster. The other SDSS cluster candidate, PLCK~G193.3$-$46.1, is also confirmed.  The \xmm\ redshift measurements and the photometric redshift \footnote{The photometric redshift  is taken from the {\hbox{\sc Photoz}} table of the SDSS DR7 galaxy catalogue.} of the Brightest Cluster  Galaxy (BCG) are fully consistent in the cases where there are matches with SDSS clusters identified by our internal algorithm. For PLCK~G210.6+17.1  $z_{\rm phot}=0.48\pm0.02$ compared to $z_{\rm Fe}=0.48\pm0.02$. For PLCK~G193.3$-$46.1, $z_{\rm phot}=0.65\pm0.06$ while $z_{\rm Fe}=0.59\pm0.02$. This supports the robustness of our SDSS analysis method and indicates that the SDSS can confirm candidates up to $z\sim0.6$, and estimate their photometric redshifts.  

%________________________________________________________________
%% Figure:SDSS
%%
\begin{figure}[t]
\center
\includegraphics[width=0.8\columnwidth, clip]{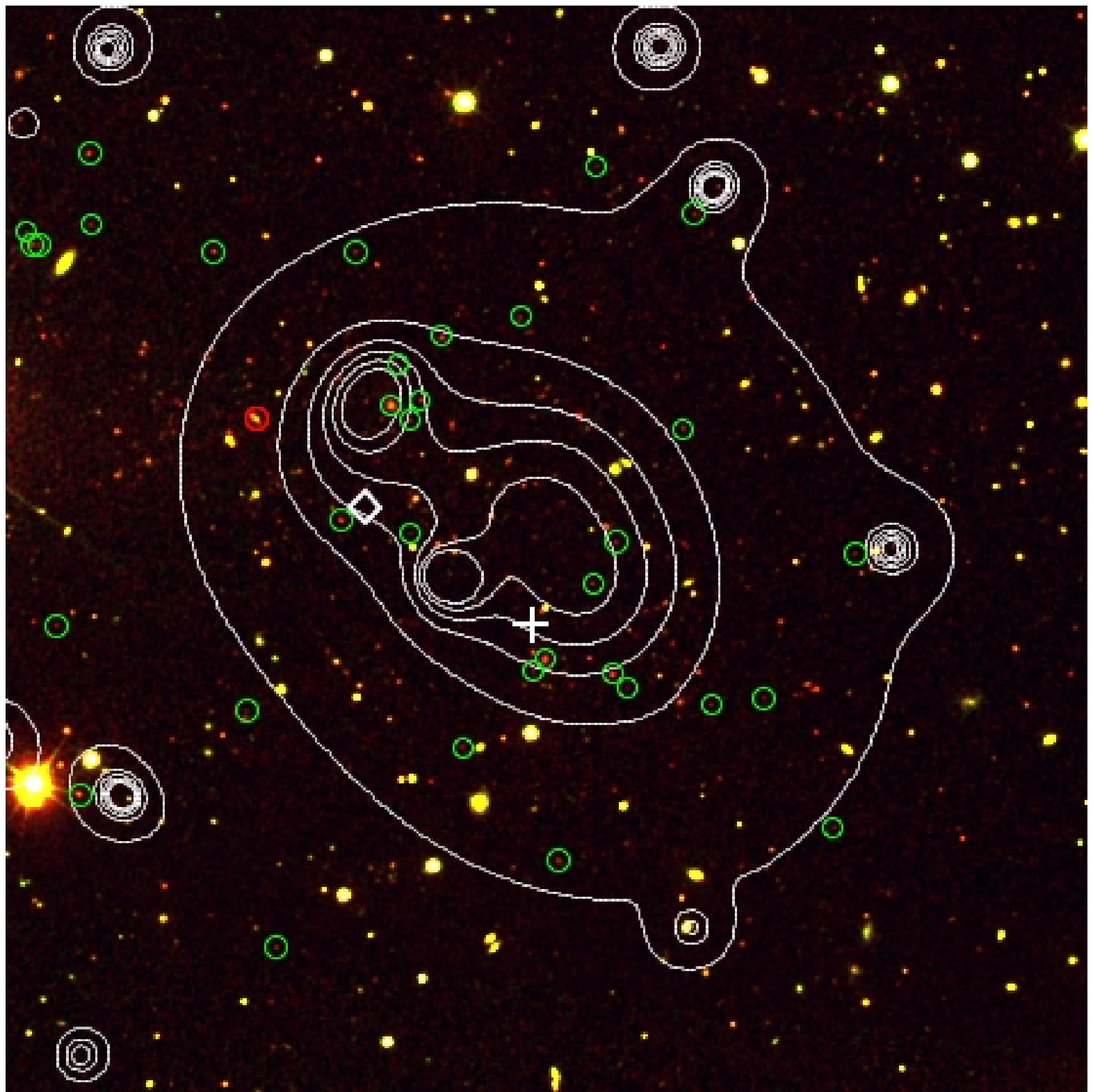}
\caption{ {\footnotesize 
SDSS colour composite image of PLCK~G193.3$-$46.1 at $z=0.6$ overlaid with  isocontours of the wavelet filterered \xmm\  image. The image size is $8\farcm9\times8\farcm9$. Green circles: cluster galaxies identified by the search algorithm. Red circle: Brightest Cluster Galaxy. Diamond: centroid of galaxy distribution.  Cross: \planck\ SZ position. 
} \label{fig:sdss}} 
\end{figure}
%________________________________________________________________

 \subsection{X-ray versus SZ properties of newly detected clusters}
\label{sec:outxsz}

%________________________________________________________________
%% Figure: Ysz-Yx and Ysz-Lx 
%%
\begin{figure*}[t]
\centering
\resizebox{0.85\hsize}{!} {
\includegraphics{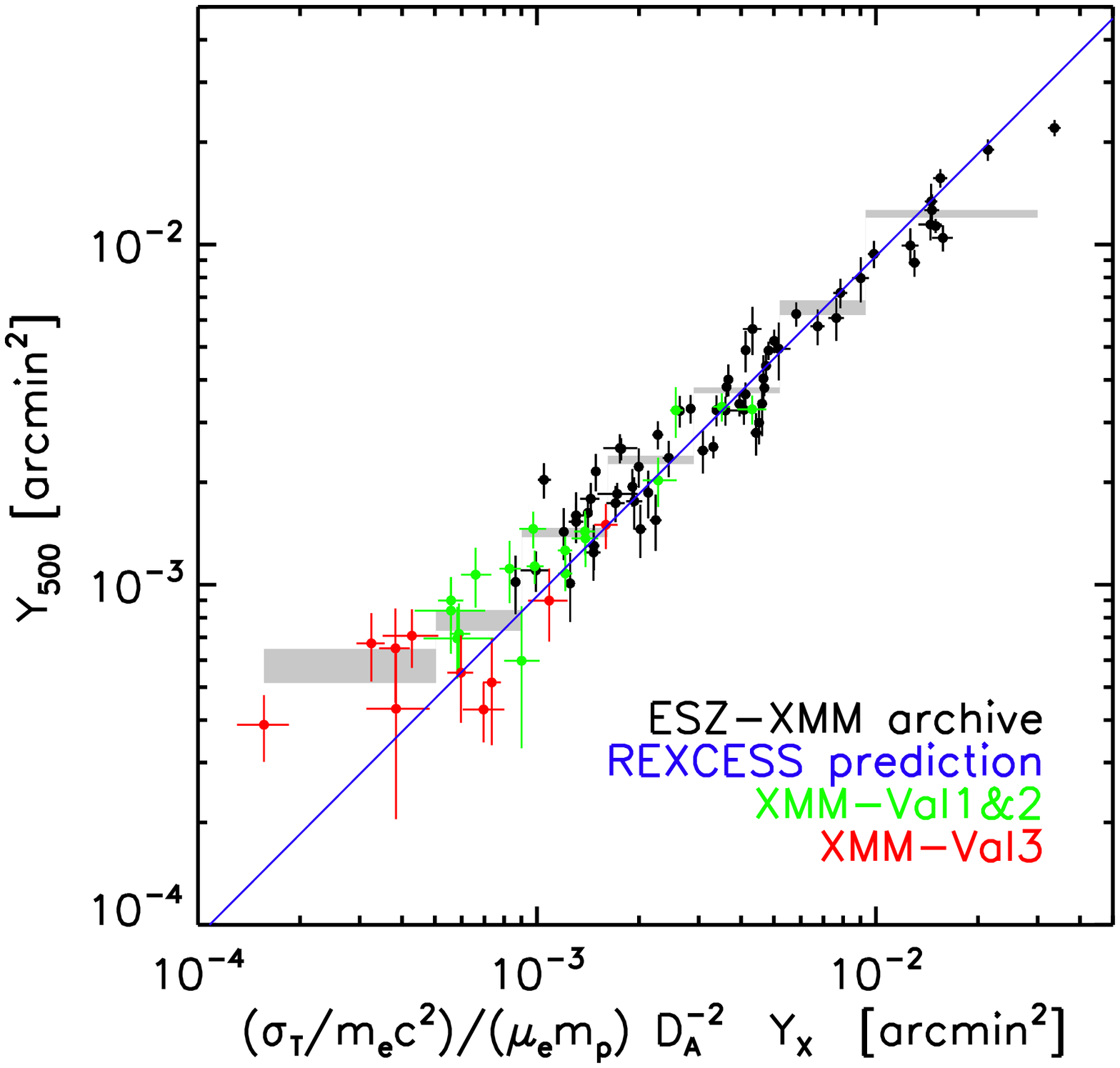}
\hspace{8mm}
\includegraphics{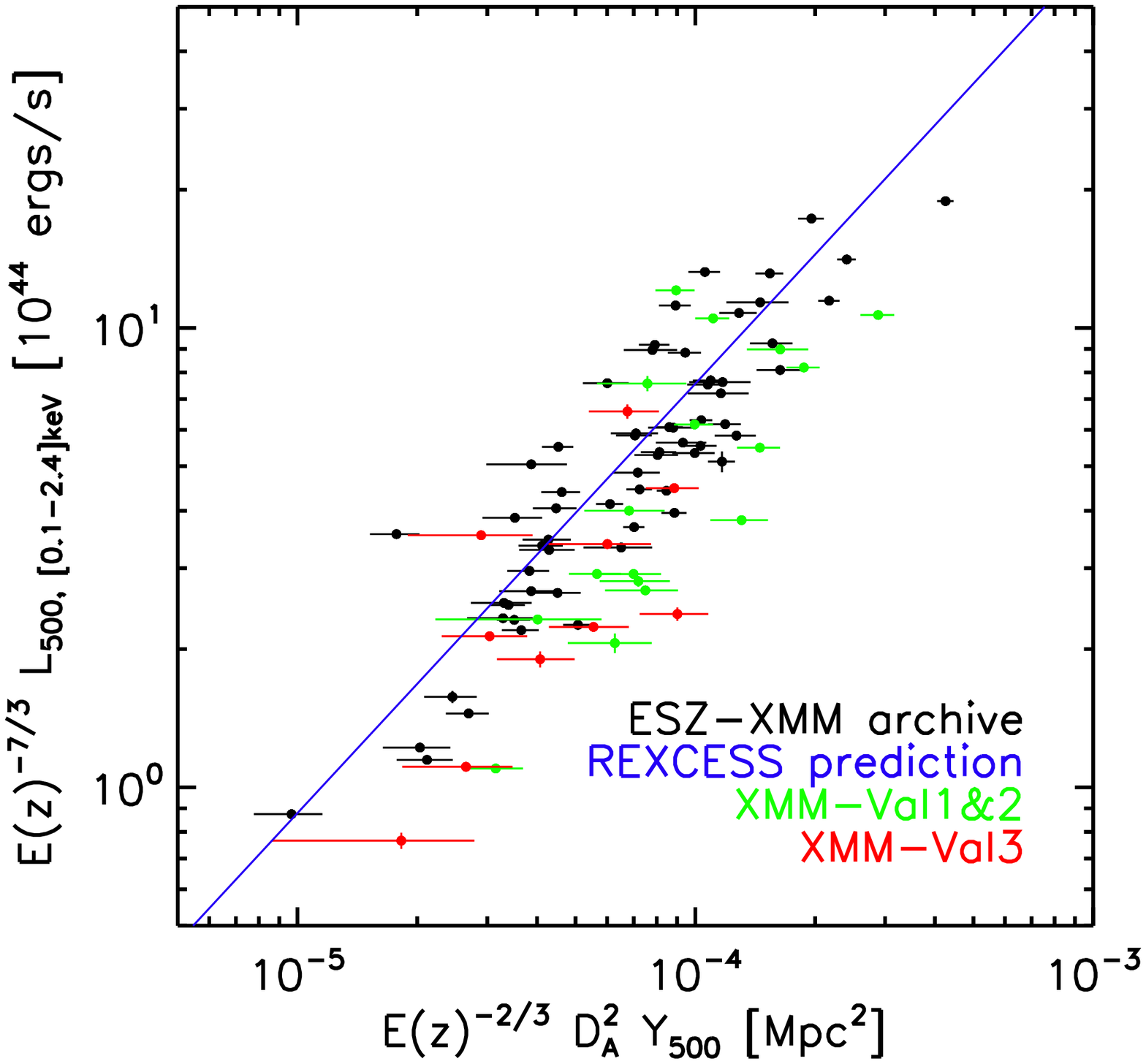}
} 
\caption{{\footnotesize 
Scaling relations for the ten new confirmed clusters (red symbols). Black points show clusters in the  \planck-ESZ sample with \xmm\ archival data as presented in \citet{planck2011-5.2b}; green points represent previously-confirmed \planck\ clusters presented in \citet{planck2011-5.1b}. The blue lines denote the $\YSZ$ scaling relations predicted from the \rexcess\ X-ray observations \citep{arn10}. {\it Left:}  Relation between apparent SZ signal ($Y_{500}$) and the corresponding normalised $\YX$ parameter.  The grey area corresponds to weighted average $\YSZ$ values in $\YX$ bins with $\pm1\sigma$ errors. {\it Right:}  Relation between X--ray luminosity and  $\YSZ$. For most data points,  uncertainties on the luminosity are smaller than the point size.} 
\label{fig:xsz}}
\end{figure*}
%________________________________________________________________

The  present study samples higher redshifts and lower SZ fluxes  than the previous \xmm\ validation observations ($0.2<z<0.97$ and $4\times 10^{-4}\,{\rm arcmin}^2<\YSZ<1.5\times 10^{-3}\,{\rm arcmin}^2$, as compared to $0.09<z<0.54$ and $6\times 10^{-4}\,{\rm arcmin}^2<\YSZ<3\times 10^{-3}\,{\rm arcmin}^2$ for the previous observations). Our previous findings, detailed in \citet{planck2011-5.1b}, are confirmed and extended to higher $z$ and/or lower $\YSZ$. The new SZ-detected clusters have, on average, lower luminosities, flatter density profiles, and a more disturbed morphology than their X-ray selected counterparts.   

The average scaled density profile (Fig.~\ref{fig:ne}) is similar to that of the {\it XMM}-Val1$\&$2 sample, and remains flatter than that of \rexcess, a representative sample of X--ray selected clusters  \citep{arn10}. The gallery of \xmm\ images (Fig.~\ref{fig:gal}) shows a variety of morphologies with three out of ten clusters exhibiting extremely flat and asymmetric/or X-ray emission. One of those is PLCK~G193.3$-$46.1 at $z=0.6$, as shown in  Fig.~\ref{fig:sdss}. Its double peaked  X--ray morphology suggests an on-going merger of two sub-clusters along the NE-SW direction, which is supported by the available SDSS data. The galaxy distribution  is not centrally peaked and  its centroid is $1\farcm9$ South/West of the BCG  position. Neither the centre of the galaxy distribution nor the BCG position coincides with any of the X--ray peaks (see Fig.~\ref{fig:sdss}). 

The new clusters follow the trends in scaling properties established from our previous follow-up (Fig.~\ref{fig:xsz}).   They are on average less luminous at a given $\YSZ$, or more massive at a given luminosity, than X-ray selected clusters.  Eight out of ten of the new clusters fall on the low luminosity side of the $L_{500}$--$\YSZ$ relation for X--ray selected clusters (Fig.~\ref{fig:xsz} right panel). As shown in the left hand panel of  Fig.~\ref{fig:xsz}, the \YSZYX\ relation for most clusters remains consistent with the \rexcess\ prediction:
\begin{equation}
\YSZ  =    0.924\,D_{\rm A}^{-2}\,C_{\rm XSZ}\,\YX, 
\label{eq:yszyx}
\end{equation} 
with $C_{\rm XSZ} =
1.416 \times 10^{-19}\,{\rm Mpc}^2/\msol\, \keV$.  However, the SZ flux levels off around $\YSZ\sim4\times 10^{-4}\,{\rm arcmin}^2$.  This turnover at low flux is clearly apparent when considering the weighted  average  $\YSZ$ values in $\YX$ bins. It deviates significantly from the prediction in the two lowest $\YX$ bins, a deviation increasing with decreasing $\YX$ (grey area in the left panel of Fig. 6). This is reminiscent of the Malmquist bias resulting from a flux cut selection. 
Due to scatter around the mean relation between the observed flux ($\YSZ$) and the `true' flux (estimated from $\YX$), objects below the flux cut are detectable but,  in order to be detected, they must be increasingly deviant from the mean relation with decreasing intrinsic flux.   The effect is more prominent  than that already observed for the {\it XMM}-Val1$\&$2 sample  \citep{planck2011-5.1b}, whereas it is negligible for the ESZ-{\it XMM}-archive sample (Fig.~\ref{fig:xsz}). This is likely due to the increasing magnitude of the Malmquist bias as a function of decreasing flux \citep[see also][]{planck2011-5.2b}.   Note that the scatter in the $\Yv$--$\YX$ relation, and thus the Malmquist bias, is likely dominated by measurements errors. $\Yv$ or $\YX$ are related to the same physical quantity, the thermal energy of the gas. The intrinsic scatter in $\Yv$ for a given $\YX$ is thus expected to be smaller  than the $<10\%$ intrinsic scatter of either  $\Yv$ or $\YX$  values at fixed mass \citep{kra06,arn07},  and thus smaller than the statistical scatter  in the $S/N\lesssim5$ regime. Outliers are present,  though, as discussed  below.

%________________________________________________________________
%% Figure: ENO
%%
\begin{figure}[bpt]
\center
\includegraphics[width=0.85\columnwidth]{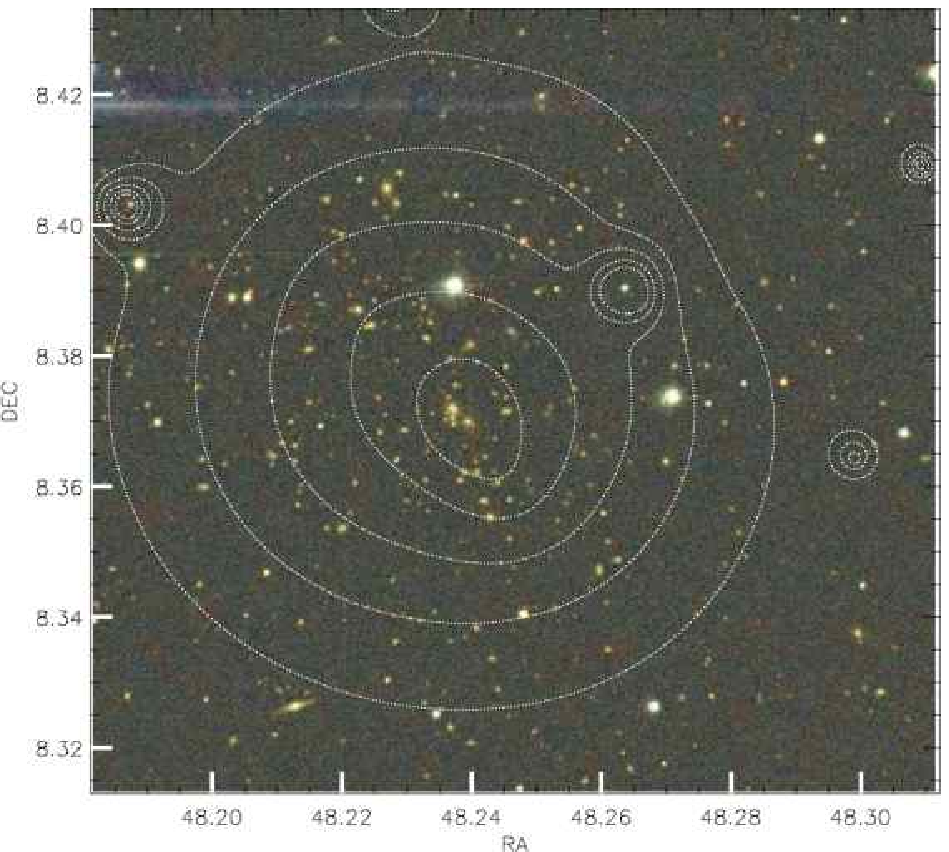}
\caption{{\footnotesize 
Colour composite images (Sloan {\it gri} filter)  of  PLCK~G171.9-40.7 observed with ENO/ IAC80 telescope. North is up, East is right and the image size is $7\farcm7\times7\farcm2$. The isocontours of the wavelet filtered \xmm\  image are overlaid. 
\label{fig:eno}}}
\end{figure}
%________________________________________________________________

The  most prominent outliers are the two lowest $D_{\rm A}^{-2}\,C_{\rm XSZ}\,\YX$ clusters,  PLCK~G235.6+23.3 and PLCK~G268.5$-$28.1, which lie 
at  $2.4\,\sigma$ and $2.8\,\sigma$, respectively,  above the expected relation (Eq.~\ref{eq:yszyx}). They could thus be due to statistical fluctuations.  However, they correspond to $\YSZ/\YX$ ratios $2.1$ and $2.5$ times higher than expected, respectively. The redshift of  PLCK~G268.5$-$28.1 is not well determined and may
 be under-estimated. We cannot thus exclude that its $\YX$ value is actually higher  (see also Sec.~\ref{sec:zxsz}).  On the other hand, PLCK~G235.6+23.3 is an unprepossessing cluster at $z=0.37$ with no remarkable X--ray properties, and for which we have very accurate SZ and X-ray measurements.   Only one such outlier in terms of $\YSZ/\YX$ ratio appears in  the  ESZ-{\it XMM}-archival sample of 62 clusters: RXCJ0043.4$-$2037,  a relaxed cluster at $z=0.29$ \citep[][]{fin05} as can be seen in Fig.~\ref{fig:xsz}. 
A complete follow-up of \planck\ candidates is required to quantify the intrinsic scatter in the \YSZYX\ relation and its associated Malmquist bias. Only then can one compare  the true dispersion in  the $\YSZ/\YX$ relation with that  established from the ESZ-{\it XMM}-archival sample.

\section{Redshift determination}

\subsection{New optical redshift determinations}
In this section we present new optical redshift determinations for ten confirmed clusters of the {\it XMM}-Val1$\&$2 sample and for two of the present sample.

\subsubsection{ENO observations}

PLCK~G171.9$-$40.7 and PLCK~G100.2$-$30.4 were observed  with the 0.82~m IAC80 telescope at the Observatorio del Teide (Tenerife, Spain) as part of a larger campaign for optical follow-up of newly detected \planck\ candidates. Images were taken in four Sloan filters, {\it griz}, with the CAMELOT camera. This camera  is equipped with a 2048$\times$2048 pixel CCD (0.304 arcsec per pixel), resulting in a field of view of $10\farcm4 \times 10\farcm4$.

The data reduction included all standard calibrations, i.e., bias and flat field corrections and astrometric calibration. Source detection was undertaken by running \textsc{SExtractor} \citep{ber96} on the {\it i}-band images, and photometry on all bands was obtained in double-image mode. For source detection we used a detection threshold of 3$\sigma$ in the filtered maps, which corresponds to a S/N$\sim$ 6. All sources classified as stellar objects, based on a stellarity index greater than 0.8 in all bands (given by SExtractor) were excluded from our sample. We applied galactic extinction correction based on the dust maps by \citet{sch98}. 

%________________________________________________________________
%% Figure: ESO
%%
\begin{figure}[bpt]
\center
\includegraphics[width=0.85\columnwidth]{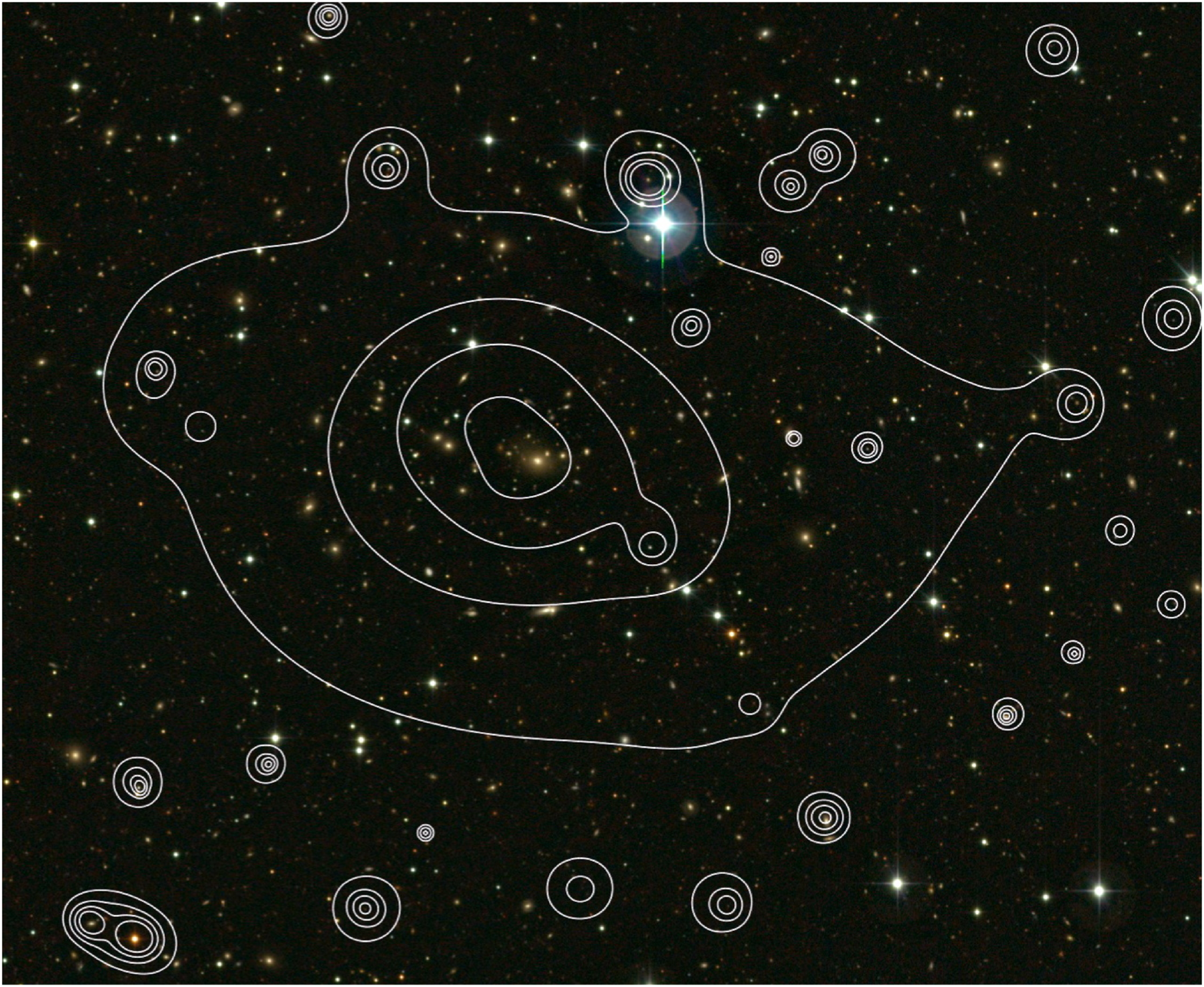}
\caption{{\footnotesize 
 VRI colour composite image of PLCK~G262.2$+$34.5 observed  with the ESO/MPG 2.2m telescope, with exposure times of 0.5h, 1.4h, and 0.5h in the V, R, and I-band, respectively.The isocontours of the wavelet filterered \xmm\  image are overlaid. 
North is up and east to the left, and the image size is $13\farcm6 \times 11\farcm1$.  Although the X-ray morphology of PLCK~G262.2$+$34.5 is flat rather than centrally peaked, the X-ray centre coincides well with the location of the BCG. 
The field also contains a large number of X-ray point sources with optical counterparts. 
\label{fig:eso}}}
\end{figure}
%________________________________________________________________

PLCK~G171.9$-$40.7 was observed in each Sloan {\it griz}  filter in 4000~s exposures. The limiting magnitudes reached are 
23.2, 21.1, 20.6 and 20.6 magnitudes for $g,r,i,z$, respectively. The colour composite image in Fig.~\ref{fig:eno} clearly shows a galaxy overdensity coincident with the X--ray image. The  BCG is only slightly offset from the  X--ray peak. The final catalogue contains 384 sources, for which we obtained photometric redshifts using the BPZ code \citep{ben00}. 
The redshift estimate for each individual galaxy is based on all four  filters, and is obtained by fitting a set of SED (spectral energy 
distribution) templates \citep[see details in][]{ben00}. The BPZ code provides the Bayesian posterior probability distribution
function (pdf) for the redshift of each object. We have calibrated the code for our set of four filters using a subsample of 5000 galaxies from SDSS DR8 with spectroscopic redshift, $z_{\rm spec,SDSS}$.  The standard deviation of the difference between $z_{\rm spec,SDSS}$ and the  photometric redshift, $z_{\rm phot,BPZ}$, obtained applying the BPZ  code to this subsample,   is  $\Delta z = 0.03$. The deviation is  similar  for the whole sample and  for the two different redshift intervals, $0<z<0.2$ and $0.2<z<0.4$.  In a conservative approach, we used this deviation as systematic uncertainty on  cluster redshift. The statistical uncertainty is negligible in comparison.   For PLCK G171.9-40.7, we use 29  cluster members to infer the photometric redshift, and we obtain  $z_{\rm phot}=0.31 \pm 0.03$. 

The data taken for PLCK~G100.2$-$30.4 were already presented in \citet{planck2011-5.1b}. Images have accumulated integration times of 3000\,s  in each filter and limiting magnitudes of 22.9, 21.7, 20.1 and 20.2 magnitudes for $g,r,i,z$, respectively. Reduction and catalogue compilation followed the same steps as detailed above for PLCK~G171.0$-$40.7. With respect to the results presented in \citet{planck2011-5.1b}, the main improvement is that the final images were photometrically re-calibrated using galaxies from SDSS DR8. The initial catalogue contains 452 sources for which photometric redshifts were derived.  The object has a photometric redshift of $z_{\rm phot}=0.34 \pm 0.03$, estimated from the 72 identified cluster members.

%________________________________________________________________
\begin{table}[bpt]
\begingroup
\caption{ {\footnotesize Optical redshift  data for \xmm\ confirmed clusters. The references are given in Column (4), with (s) denoting  an optical spectral redshift and (p) a photometric redshift.  The redshift from \xmm\ spectral fitting \citep[][or present work]{planck2011-5.1b} are given in Col (2) for comparison.}
\label{tab:zcomp}}
\begin{minipage}{0.8\columnwidth}
\center
\begin{tabular}{lllll}
\toprule
\multicolumn{1}{l}{Name} & \multicolumn{1}{c}{$z_{\rm Fe}$} & \multicolumn{1}{c}{$z_{\rm opt}$} &\multicolumn{1}{c}{Ref.} \\
\midrule
PLCK~G100.2$-$30.4&	$0.31\pm0.03$	&$0.34\pm0.03$&1 (p) \\
PLCK~G171.9$-$40.7&	$0.27\pm0.01$	&$0.31\pm0.03$&1 (p) \\
PLCK~G193.3$-$46.1&	$0.59\pm0.02$	&$0.65\pm0.05$&2 (p)  \\
PLCK~G205.0$-$63.0&	$0.31\pm0.01$	&$0.31\pm0.02$&3 (p) \\ 
PLCK~G210.6+17.1&	$0.48\pm0.02$	&$0.478\pm0.01$&2 (p)  \\
PLCK~G214.6+37.0&	$0.45\pm0.02$&$0.44\pm0.02$&3 (p)  \\
PLCK~G241.2$-$28.7&	$0.42\pm0.01$	&$0.41\pm0.02$&3 (p)  \\
PLCK~G262.2+34.5&	$0.21\pm0.02$	&$0.23\pm0.02$&3 (p)  \\
PLCK~G262.7$-$40.9&	$0.39\pm0.01$	&$0.422 $&4 (s)\\  
PLCK~G266.6$-$27.3&	$0.94\pm0.02$	&$0.972 $&5 (s)  \\
PLCK~G271.2$-$31.0&	$0.37\pm0.005$	&$0.32\pm0.01$&5 (p) \\  
PLCK~G272.9+48.8&	$0.40\pm0.01$	&$0.46\pm0.05$&3 (p)  \\
PLCK~G277.8$-$51.7&	$0.44\pm0.02$	&$0.438 $&5 (s)  \\
PLCK~G285.0$-$23.7&	$0.39\pm0.005$	&$0.37\pm0.02$&6 (p)  \\ 
PLCK~G285.6$-$17.2&	$0.35\pm0.01$	&$0.37 \pm0.02$&3 (p)   \\
PLCK~G286.3$-$38.4&	$0.31\pm0.01$	&$0.307\pm0.003$& 6 (s) \\  
PLCK~G286.6$-$31.3&	$0.22\pm0.005$	&$0.17\pm0.02$&3 (p)  \\
PLCK~G287.0+32.9&	$0.39\pm0.01$	&$0.37\pm0.02$&3 (p)  \\
PLCK~G292.5+22.0&	$0.31\pm0.02$	&$0.29\pm0.02$&3 (p)  \\
PLCK~G334.8$-$38.0&	$0.35\pm0.03 $	&$0.37\pm0.02$&3 (p)  \\
\bottomrule
\end{tabular}
\end{minipage}
%}
\\
{\footnotesize References: 
$(1)$ Present work from ENO/IAC80 observations;
$(2)$ SDSS-DR7 data base http://www.sdss.org/dr7/;
$(3)$  Present work from ESO/MPG2.2m observations;
$(4)$ \citet{sif12} ACT~J0438$-$5419
$(5)$ \citet{wil11}; SPT-CLJ0615-5746, SPT-CLJ0549-6204, SPT-CLJ0254-5856, respectively.
$(6)$ \citet{planck2011-5.2b}
}
\endPlancktable 
 \endgroup
\end{table}
%________________________________________________________________

\subsubsection{ESO observations}
\label{sec:eso}

Optical imaging observations of the \xmm\ confirmed clusters discussed in \citet{planck2011-5.1b} were also carried out on the ESO/MPG 2.2m telescope at La Silla Observatory using the Wide-Field Imager (WFI), which has a field of view of $33\arcmin\times 34\arcmin$ and pixel size $0\farcs238$. Each cluster was observed in the $V$, $R$, and $I$-bands in typical seeing conditions of $1.0$--$1.2\arcsec$, for total exposure times of at least 0.5h (consisting of $5\times 360$s dithered sub-exposures) per filter. The raw data were calibrated using standard techniques and individual exposures were re-registered and combined using the USNO-B1 catalogue as an astrometric reference. As an illustration, the VRI colour composite image of PLCK~G262.2$+$34.5 is shown in Fig~\ref{fig:eso}.

Galaxies that were simultaneously identified in the combined $V$, $R$, and $I$ images were plotted in a $V-R$ vs. $R-I$ colour-colour diagram. 
For each cluster, an overdensity of red galaxies, corresponding to the early-type cluster galaxies, was identified in colour-colour space. 
Galaxies associated with this overdensity in colour-colour space and also spatially coincident (to within \hbox{$\sim5\arcmin$}) with the X-ray cluster position were assumed to be early-type cluster members. Predicted $V-R$, $V-I$ and $R-I$ colors of early-type cluster galaxies as a function of redshift were calculated by convolving the `E0' template galaxy spectrum of \citet{col98} with the combined (filter+CCD) response curves for the $V$, $R$ and $I$ filters at WFI.

%________________________________________________________________
%% Figure:zcomp
%%
\begin{figure}[bpt]
\center
\includegraphics[width=0.75\columnwidth, clip]{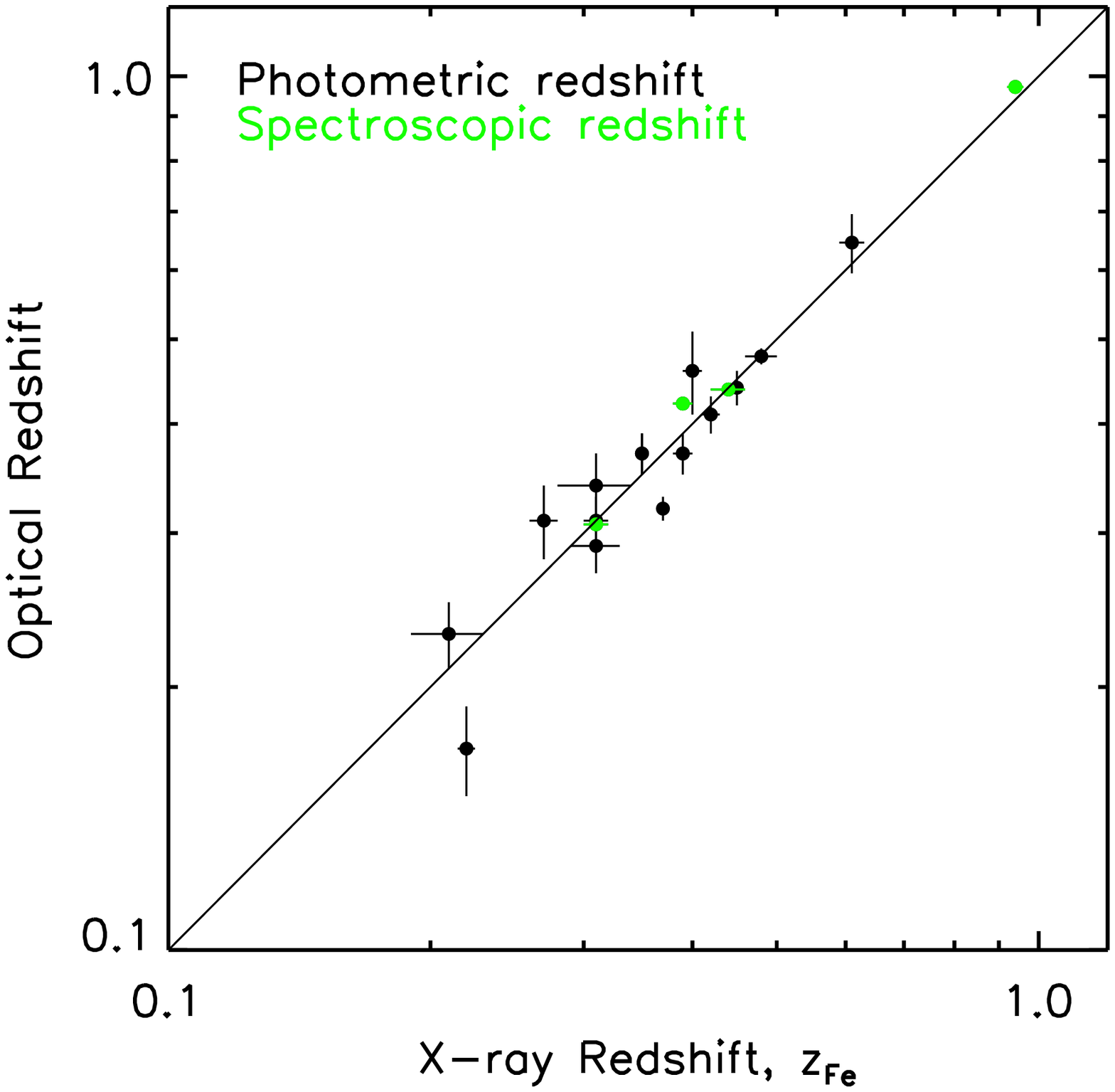}
\caption{ {\footnotesize 
Comparison between the redshift estimated from optical data and that from \xmm\ spectroscopy. 
} \label{fig:zcomp}} 
\end{figure}
%________________________________________________________________

%________________________________________________________________
%% Figure: spectres
%%%
\begin{figure*}[t]
\center
\resizebox{0.7\hsize}{!} {
\includegraphics[height=3cm]{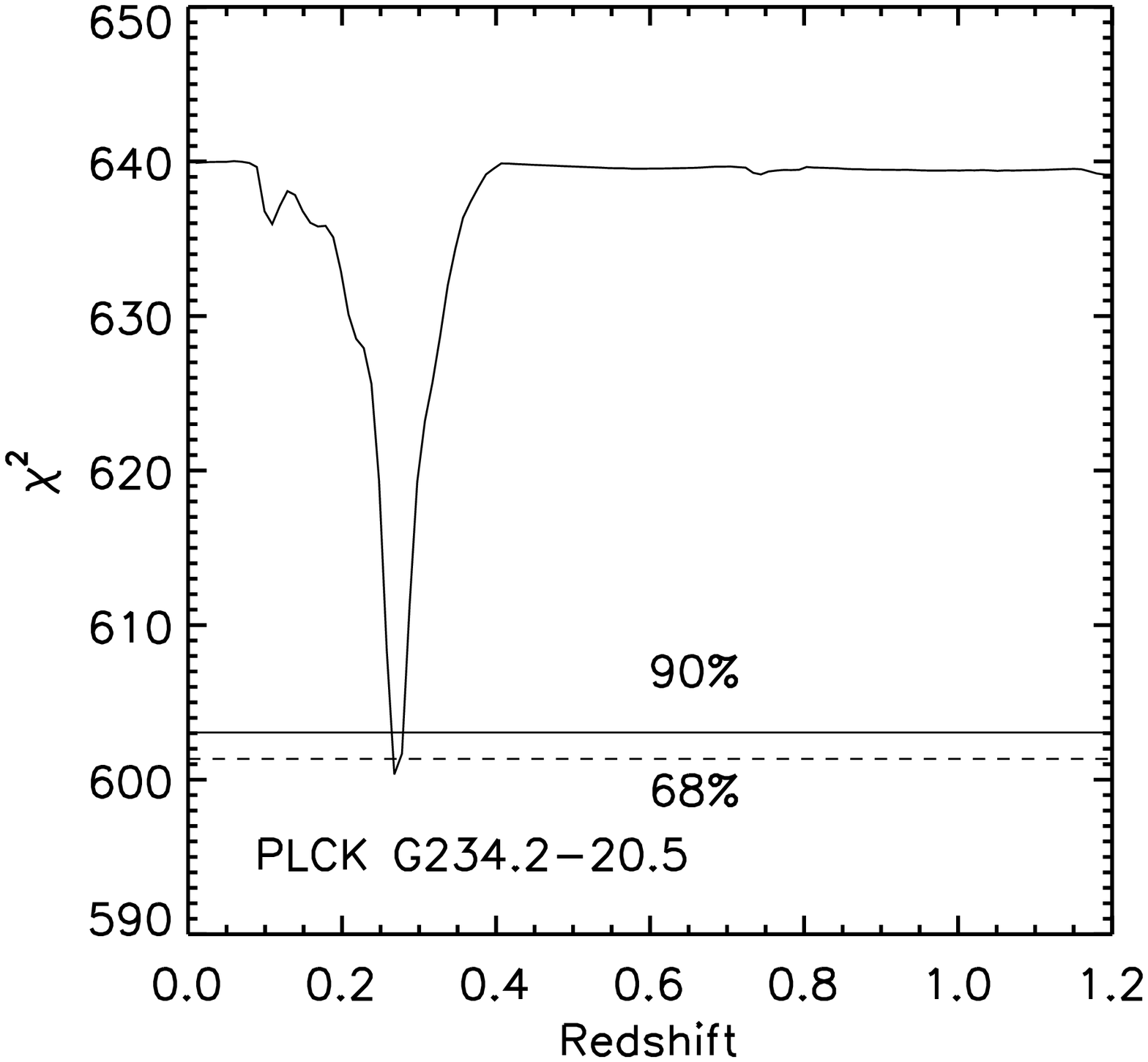}
\hspace{4mm}
\includegraphics[width=3cm,angle=270,origin=br,keepaspectratio]{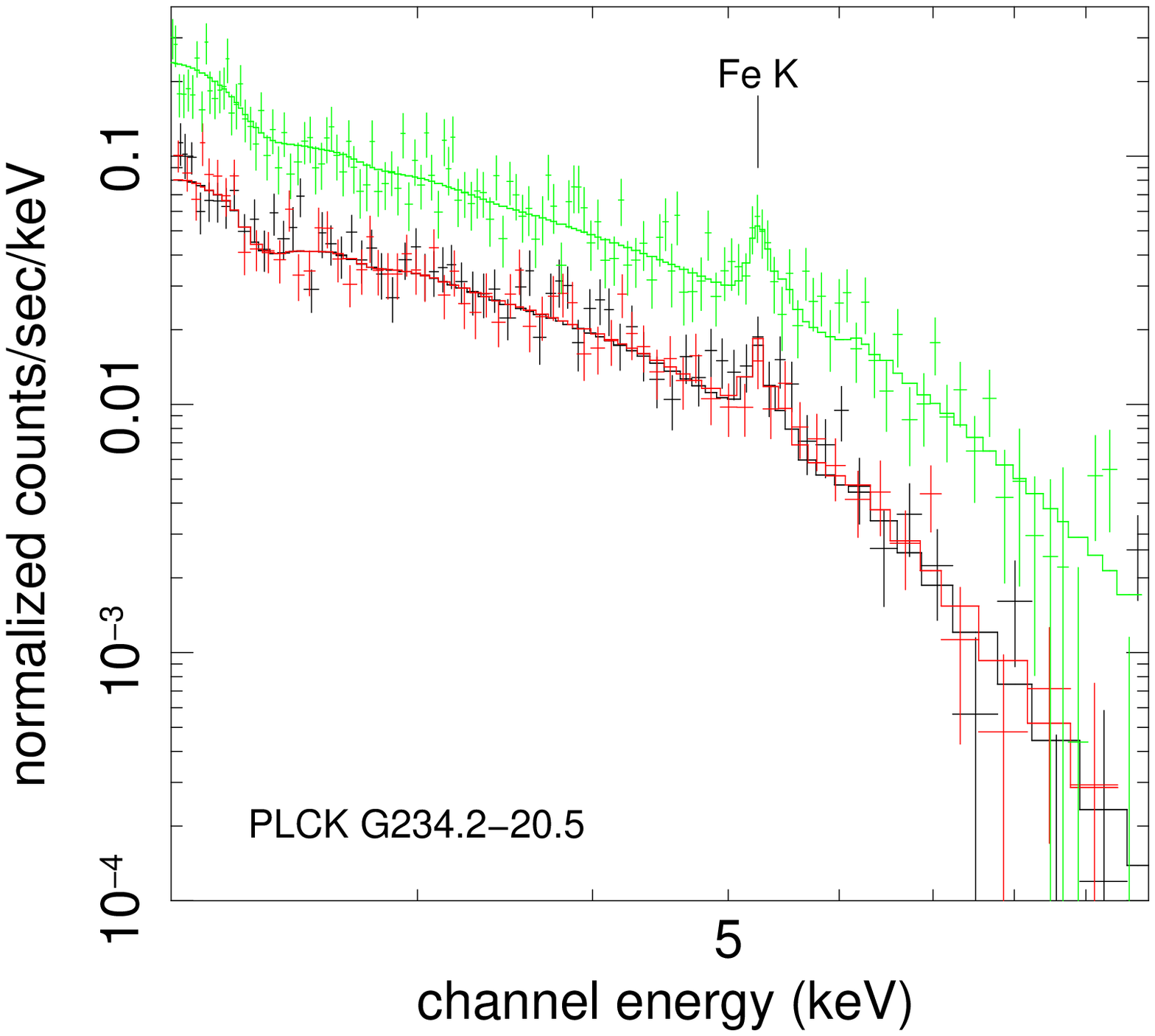}
} 
\\[5mm]
\resizebox{0.7\hsize}{!} {
\includegraphics[height=3cm]{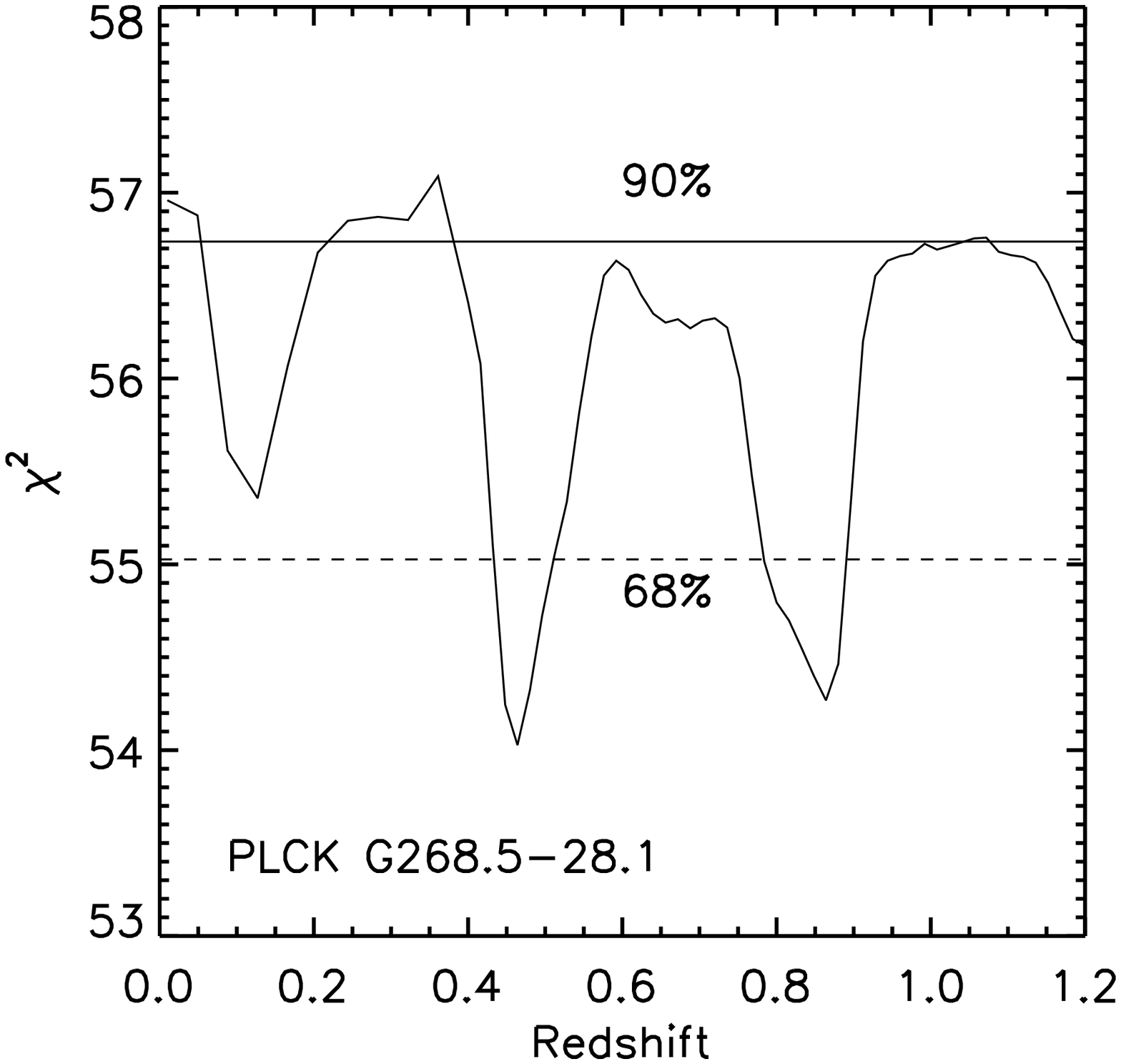}
\hspace{4mm}
\includegraphics[width=3cm,angle=270,origin=br,keepaspectratio]{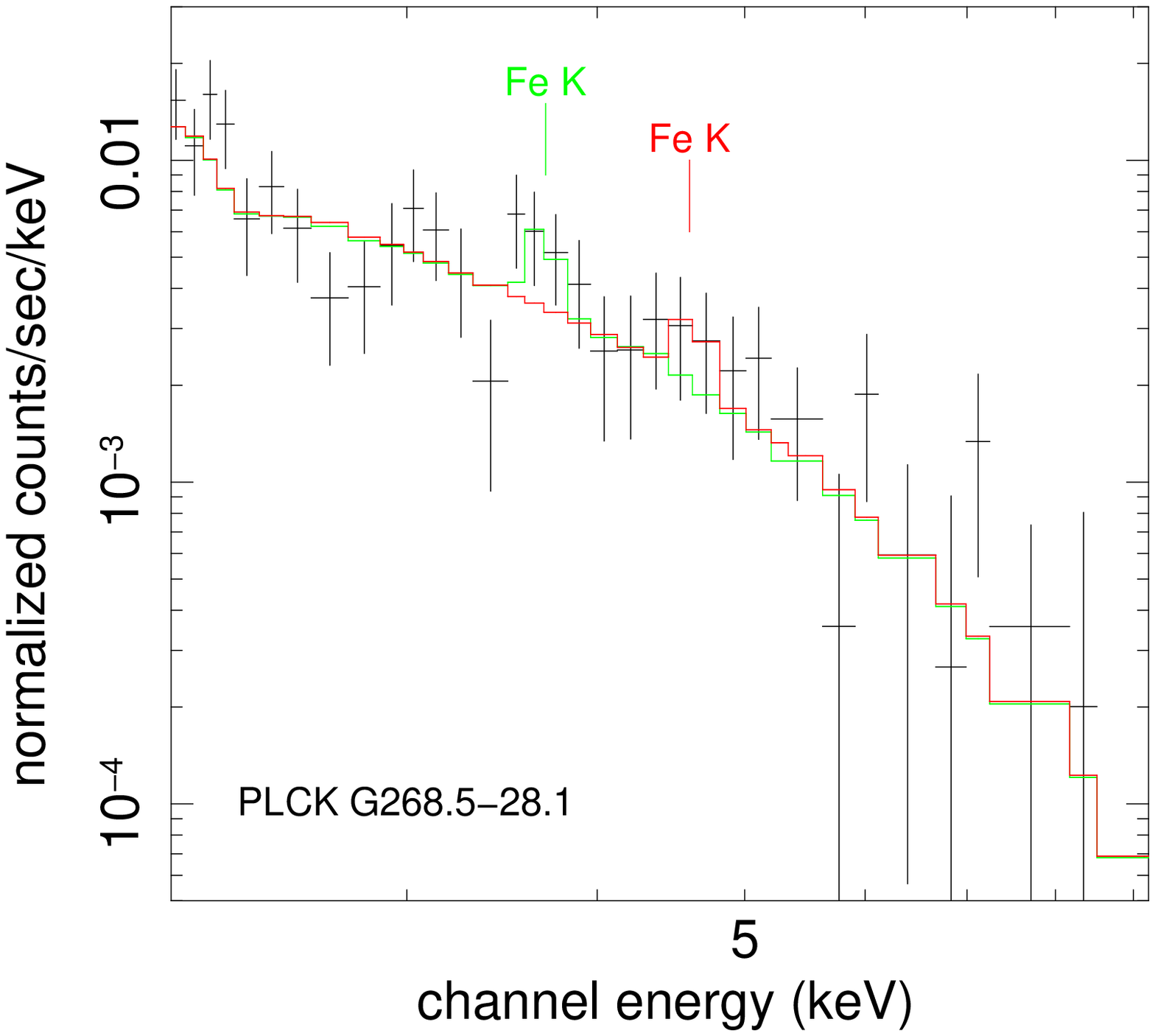}
} 
\caption{{\footnotesize 
Redshift determination from \xmm\ spectroscopy for the highest quality data ({\it top row}, PLCK~G234.2$-$20.5) and lowest quality data  ({\it bottom row}, PLCK~G268.5-28) in the sample. {\it Left panels:} Variation of $\chi^2$ with $z$ when fitting the EPIC spectra, all other parameters being let free. The dashed and full lines correspond to the $68\%$ and $90\%$ error range, respectively. {\it Right panels:} EPIC spectra (data points with errors), together with the best-fitting model thermal model (solid lines) with the position of the redshifted Fe K line marked. Only the data points above 2 keV are shown for clarity, but data down to 0.3 keV are used in the spectral fitting. For PLCK~G234.2-20.5 ({\it top right panel}), the Fe-K line complex is clearly detected in the EPIC MOS1\&2 (red and black points) and pn (green points) spectra. For PLCK~G268.5-28 ({\it bottom right panel}),
only MOS data can be used (see Sec.~3.1)  and the spectra are of poor statistical quality. The redshift estimate is ambiguous and the  $\chi^2$ distribution ({\it bottom left panel}) shows several minima. The MOS1\&2 spectra, summed  for clarity, are compared to the best fitting model for  $z=0.47$ (red line) and $z=0.86$ (green line), corresponding to the two lowest minima. } \label{fig:specz}}
\end{figure*}
%________________________________________________________________

A photometric redshift estimate  was then derived by comparing the median $V-R$, $V-I$, and $R-I$ colors of the early-type cluster galaxies to these predictions and averaging the three resulting redshift values.  We estimated how typical fluctuations in the photometric zero-point throughout the night translate into uncertainties in the measured $V-I$, $V-R$ , $R-I$ colors of galaxies. Given the predicted relation between these colors and the redshift of early-type galaxies, the estimated  $1\sigma$ redshift accuracy is $\Delta z = 0.02$. 
The new photometric redshift estimates for ten clusters observed with WFI are given in Table~\ref{tab:zcomp}. They were derived from at least 70 photometic redshifts per cluster (mean number of 120).

 \subsection{Comparison between optical and X--ray $z$ estimates}

Optical redshifts for twenty  \xmm\ confirmed \planck\ clusters are now available. This includes the fourteen measurements presented here or in \citet{planck2011-5.1b}, values from the literature for the four clusters discovered independently by ACT or SPT \citep{mar11,wil11}, and two photometric redshifts that we retrieved from SDSS data.  The values and references are given in Table~\ref{tab:zcomp},  together with \xmm\ derived value from the X--ray spectra.  For clusters with ambiguous X--ray redshift estimates ($Q_{\rm SZ}<2$), the values\footnote{The other possible $z_{Fe}$ values are given in the footnote of Table~\ref{tab:xray}.} refer to the most significant  $\chi^2$ minimum used above to calculate physical properties.The optical and X--ray estimates are compared  in Fig.~\ref{fig:zcomp}. The agreement is excellent, with a weighted mean ratio of $1.002$ and a standard deviation around equality of $0.08$. The  X--ray and optical spectroscopic redshifts (three  clusters) are consistent within  $\Delta(z)<0.02$.

%________________________________________________________________
%% Figure: z from Ysz-Yx and Ysz-Lx 
%%
\begin{figure*}[t]
\center
\resizebox{0.85\hsize}{!} {
\includegraphics{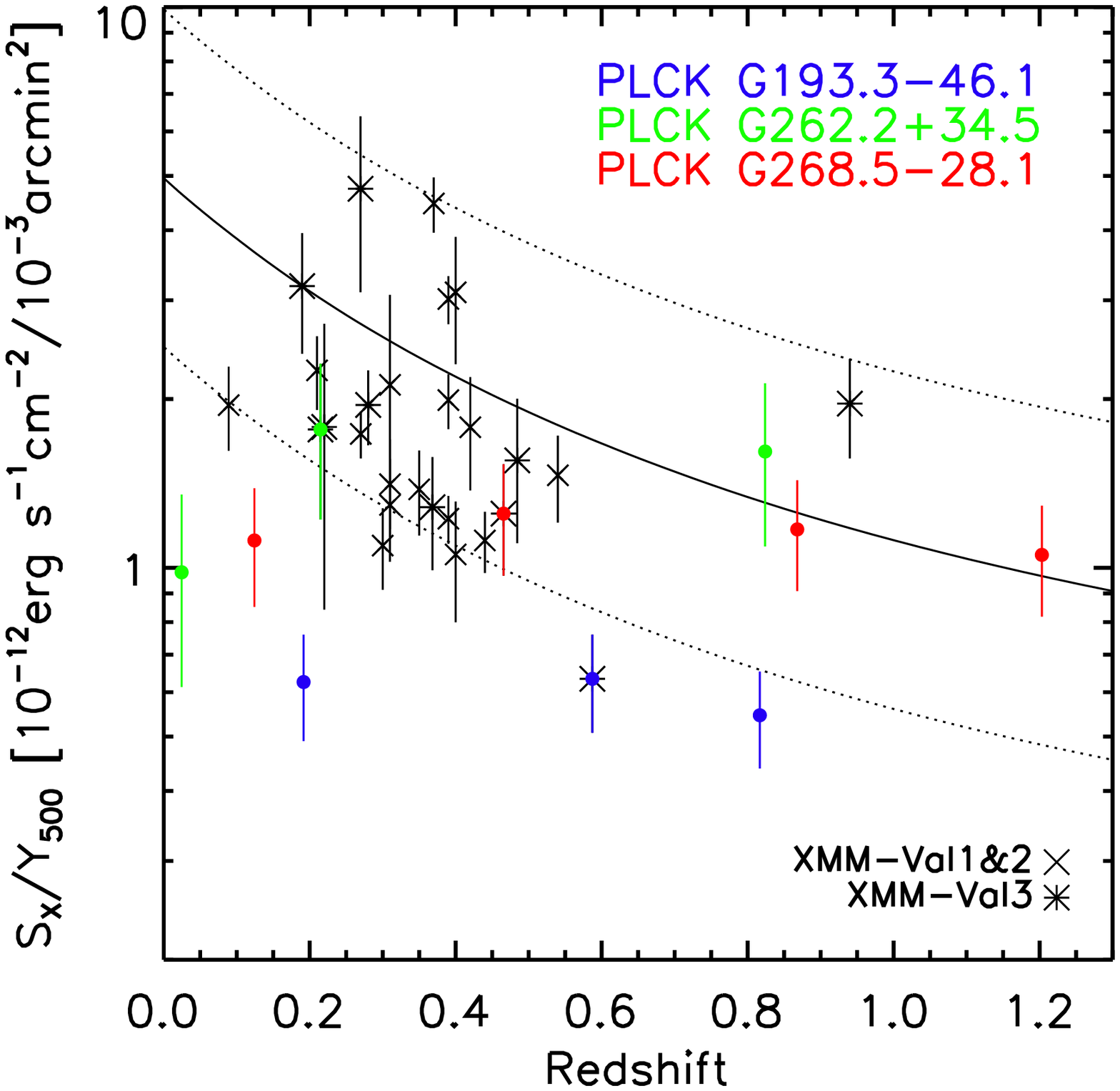}
\hspace{8mm}
\includegraphics{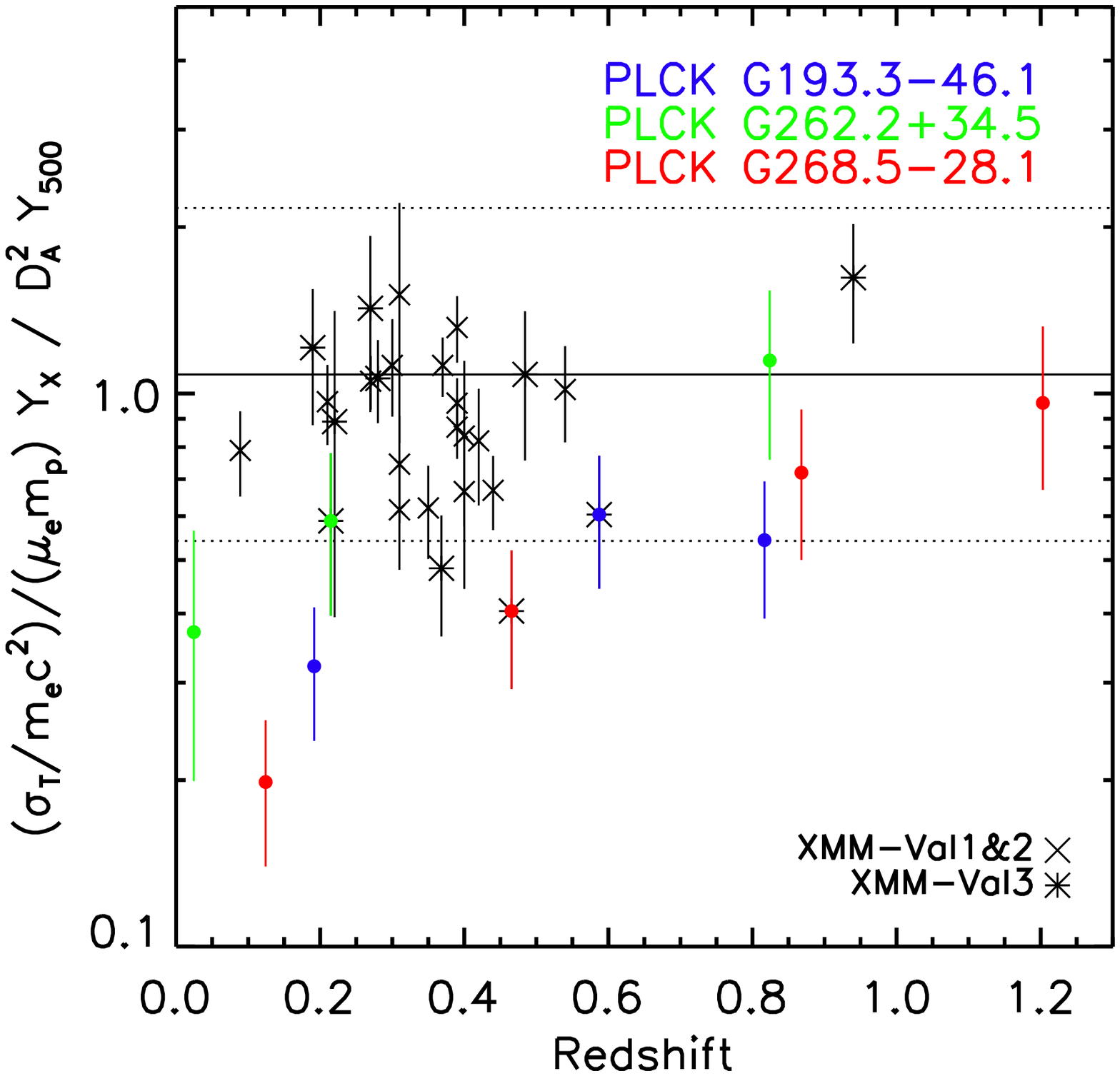}
}
\caption{{\footnotesize  Variation with redshift of the ratio between the X-ray and SZ flux ({\it left panel}) and between $\YX$ and $D_{\rm A}^2\YSZ$ ({\it right panel}).  Line: locus established from scaling relations \citep{planck2011-5.2b,arn10}. The dotted lines correspond to a factor of two  above or below the mean relation. Black points:  data for new \planck\ candidates confirmed with \xmm. Colour points:  data for clusters with ambiguous X--ray redshift estimates, one colour per cluster. Each point corresponds to one of the redshift solutions for an individual cluster, as derived from the $\chi^2(z)$ minima (see Fig.~\ref{fig:specz}).  
\label{fig:zxsz}}}
\end{figure*}
%________________________________________________________________

\subsection{Redshift estimate from a combined X--ray and SZ study}
\label{sec:zxsz}

For three clusters in the present sample, PLCK~G193.3$-$46.1, PLCK-G262.2+34.5 and PLCK~G268.5-28.1, the X--ray $z$ estimates are ambiguous ($Q_{\rm z}=1$) and the spectral fit as a function of $z$ exhibits several $\chi^2$ minima that cannot be distinguished at the $68\%$ confidence level, as illustrated in Fig.~\ref{fig:specz} (Bottom left panel). This arises when the Fe-K line complex is detected at low significance and statistical fluctuations in the spectra  of the same magnitude can  mimic the presence of a line (see the bottom right panel of Fig.~\ref{fig:specz}). Low statistical quality data arises  because the cluster is intrinsically faint or  the X-ray observations are affected by high background conditions. For comparison, the top row of Fig.~\ref{fig:specz}  shows the results for PLCK-G234.2-20.5, for which the data quality are the highest in the sample.

Optical follow-up observations are obviously required to obtain a precise redshift. However, better X--ray redshift estimates are useful for optimising any potential follow-up, e.g., for the use of the most appropriate optical facility or for deciding the pertinence of deeper X--ray follow-up based on known physical properties.
In principle, the redshift can be constrained by combining X-ray and SZ data, following a method similar to that used historically to constrain the Hubble constant.  The method relies on the different distance dependence of the X-ray and SZ measurements.
Here we examined the redshift constraining power of both the \YXYSZ\ and the \LXYSZ\ relations, using the relations established by \citet{planck2011-5.2b} from ESZ clusters with archival \xmm\ data.  We consider the three clusters with ambiguous X--ray redshift estimates, including the two clusters, PLCK~G193.3$-$46.1 and PLCK-G262.2+34.5, for which a photometric redshift is available (Table 3). Use of the latter allows us to undertake an internal consistency check.

The X--ray luminosity in the $[0.1$--$2.4]$\,\keV\ energy band scales quasi-linearly with $D_{\rm A}^2\YSZ$. Using the normalisation of  the $L_{\rm X}$--$D_{\rm A}^2\YSZ$ relation and its $z$ dependence, given in Table~2 of \citet[][]{planck2011-5.2b}, and taking into account the $z$ dependence of the luminosity-distance, one can write:
\begin{equation}
\frac{\left[F_{\rm X}/{\rm 10^{-12}\, erg\,s^{-1}\,cm^2}\right]} {\left[\YSZ/{\rm 10^{-3}\,arcmin^2}\right]}=   4.95\, E(z)^{5/3}\, (1+z)^{-4}\, K(z)
\label{eq:fluxxsz}
\end{equation}
where $F_{\rm X}$ is the X--ray flux at Earth in the same band and $K(z)$ is the K correction. The K correction increases with $z$,  with a typical value of $K=1.24$ at $z=0.5$ for a $\kT\,=\,6\,\keV$ cluster. We can neglect the temperature dependence of the K correction, which is much smaller than the typical dispersion of the \LXYSZ\ relation for the energy band and mass range under consideration.

The theoretical relation is plotted in the left hand panel of Fig.~\ref{fig:zxsz}. For each cluster, we then estimated the X-ray flux and $\YSZ$, fixing $z$ to each possible value in turn. The flux estimates depend on physical cluster parameters such as size $\theta_{500}$ and temperature, whose estimate  depends in turn on $z$ and requires data of sufficient quality. As can be seen in the Figure, in practice the measured flux ratio depends weakly on the assumed $z$. This simply reflects the fact that the fluxes are the quantities most directly related to the raw measurements. If data are of insufficient quality, the ratio can also simply be estimated at a fiducial $z$ and $\kT$.  More importantly, since the true ratio depends on $z$, this redshift can be constrained from the measured ratio and Eq.~\ref{eq:fluxxsz}. Unfortunately, as can be seen in Fig.~\ref{fig:zxsz}, the large dispersion around the relation limits the constraints one can achieve with this method. The variation beyond  $z\sim0.2$ is no more than a factor of two, meaning that one cannot distinguish between a factor of two under-luminous outlier at $z=0.2$ and  a `normal' cluster at $z=1$.  The lack of constraining power is exacerbated by the very nature of  the clusters in question; those with poor $z$ estimates are generally objects with low intrinsic X-ray fluxes. 

We thus also examined the  \YXYSZ\ relation, which exhibits a lower intrinsic scatter. \citet{planck2011-5.2b} showed that the \YXYSZ\ relation is consistent with that derived from \rexcess.
The $C_{\rm XSZ}\,\YX/(D_{\rm A}^2\YSZ)$  ratio is fixed from Eq.~\ref{eq:yszyx}, while its estimate from X-ray and SZ data depends on $z$. This dependence is complex and does not follow a simple analytical law.  For each assumed $z$ in turn, the parameters must be derived from  X-ray data and SZ data  re-processing. The estimated ratio increases with the assumed $z$, as illustrated in the right-hand panel of Fig.~\ref{fig:zxsz}.  Although its dispersion is smaller, the \YXYSZ\  relation does not provide better constraints. For newly detected clusters, this method is limited by  1) the large statistical uncertainty of SZ data and 2) the possibility that the cluster is an outlier (this latter being all the more important because of the Malmquist bias).  This is perfectly illustrated in the case of PLCK\,G268.5$-$28.5. A redshift as low as $z=0.15$ is very unlikely. However, the cluster could either be at $z=1.2$ if it perfectly follows the mean relations, or it could be an  under-luminous, low $\YX$ cluster  at $z=0.47$ (the best X--ray estimate). The cases of PLCK\,G193.3$-$46.1 and  PLCK\,G262.2+34.5 are  very similar: only the lowest $z$ solution can be excluded.  On the other hand, the redshifts indicated by optical data,  $z_{\rm phot}\sim0.60$ and  $z_{\rm phot}\sim0.23$, respectively, are indeed allowed by the present analysis. However, higher $z$ solutions yield X/SZ values closest to the theoretical relations. This again illustrates the limitation of the method  in the presence of scatter.  

In summary, we find that the use of the $Y_{\rm X}$ vs. $Y_{\rm SZ}$ and X-ray flux $F_{\rm X}$ vs. $Y_{\rm SZ}$ relations allows us to put lower limits on cluster redshifts. 

\section{Conclusion}

We have presented a further eleven \xmm\ X-ray observations of \planck\ cluster candidates, undertaken in the framework of a DDT validation programme. The sample was chosen from blind detections in all-sky maps from the first ten months of the survey and probes lower signal-to-noise and SZ quality criteria than published previously \citep{planck2011-5.2b}. Ten of the candidates are confirmed to be {\it bona fide} clusters, all of which fall below the RASS X-ray flux limit. The objects lie at redshifts $0.22 < z < 0.94$ and have masses (estimated from the $\Mv$--$\YX$ relation) in the range $(2.7\pm0.2)\times 10^{14}\,{\rm M}_\odot < \Mv < (7.8\pm0.6)\times 10^{14}\,{\rm M}_\odot$.  We detect a first indication for Malmquist bias in the $Y_{\rm SZ}$--$Y_{\rm X}$ relation, with a turnover at $Y_{\rm SZ} \sim 4\times10^{-4}$ arcmin$^2$.

This validation run clearly demonstrates the capability of the  \planck\ survey to detect clusters of a wide range of masses up to high $z$, although with a mass detection threshold that increases with redshift. We emphasise that the present sample is neither complete not representative, being  constructed to sample various SZ quality flags. While it is not {\it a priori} biased towards any specific type of cluster, it cannot be used to infer any statistical information on the parent catalogue, such as its underlying purity, or for quantifying the Malmquist bias.

We studied the pertinence of our internal quality grades assigned to the SZ detection, based on visual inspection of the reconstructed 2D SZ maps and SZ spectrum.  The single false candidate has a relatively high $S/N\sim5$, but the lowest SZ quality grade. This confirms that the quality of the \planck\ SZ detection cannot be reduced to a single global ${\rm S/N}$ and is an indication of the pertinence of our internal quality grade definition.  On the other hand, real clusters do have C grade detections. Such a grade is clearly not sufficient to exclude a given candidate. However, A and B grade detections are strong indications for a real cluster. 

We presented new optical redshift determinations of candidates previously with \xmm, obtained with ENO and ESO  telescopes.  The X-ray and optical redshifts for a total of 20 clusters are found to be in excellent agreement. We also show that useful lower limits can be put on cluster redshifts using X-ray data alone via the $Y_{\rm X}$ vs $Y_{\rm SZ}$ and X-ray flux $F_{\rm X}$ vs $Y_{\rm SZ}$ relations.

In terms of physical properties, the present clusters are similar to the first new \planck\ SZ detections presented in \citet{planck2011-5.2b}, except at lower $\Yv$ and higher mean redshift. The majority show signs of significant morphological disturbance, which is reflected in their flatter density profiles compared to those of X--ray selected systems.They are, on average, under-luminous for their mass as compared to X-ray selected clusters. 

In future work, we will explore even lower ${\rm S/N}$ detections and discuss  information from ancillary data, such as that available from SDSS or RASS, as an indicator of candidate validity.

\begin{acknowledgements}
  The \planck\ Collaboration thanks Norbert Schartel for his
  support of the validation process and for granting discretionary time for the observation of \planck\ cluster candidates.  
  The present work is based 
  on observations obtained with \xmm, an ESA science mission with
  instruments and contributions directly funded by ESA Member States
  and the USA (NASA), on observations made with the IAC80 telescope operated on the
island of Tenerife by the Instituto de Astrof'sica de Canarias in the
Spanish Observatorio del Teide and on observations collected using the ESO/MPG 2.2m telescope on La Silla under MPG programs 086.A-9001 and 087.A-9003. This research has made use of the following databases: SIMBAD, operated at the CDS, Strasbourg, France; the NED database, which is operated by the Jet Propulsion Laboratory, California Institute of Technology, under contract with the National Aeronautics and Space Administration; BAX, which is operated by the Laboratoire dÕAstrophysique de Tarbes-Toulouse (LATT), under contract with the Centre National dÕEtudes Spatiales (CNES); and the  SZ repository operated by
IAS Data and Operation centre (IDOC) under contract with CNES.
A description of the Planck Collaboration and a
list of its members, indicating which technical or scientific
activities they have been involved in, can be found at  http://www.rssd.esa.int/Planck\_Collaboration.
The Planck Collaboration acknowledges the support of: ESA; CNES and CNRS/INSU-IN2P3-INP (France); ASI, CNR, and INAF (Italy); NASA and DoE (USA); STFC and UKSA (UK); CSIC, MICINN and JA (Spain); Tekes, AoF and CSC (Finland); DLR and MPG (Germany); CSA (Canada); DTU Space (Denmark); SER/SSO (Switzerland); RCN (Norway); SFI (Ireland); FCT/MCTES (Portugal); and DEISA (EU)

\end{acknowledgements}

%\bibliographystyle{aa}
%\bibliography{PIP_22a.bib,Planck_bib.bib}

\appendix

\raggedright

\listofobjects
\end{document}

%% file: PIP_22a_authors_and_institutes.tex
%This author list corresponds to \title{Author list for PIP 22a, Proj. Ref. 5.1/5.2/5.5: Planck intermediate results. I. Further validation of new Planck clusters with XMM-Newton}
%Prepared by R. Leonardi (rleonardi@sciops.esa.int), ESAC/ESA
%This version is from Thu May 03 08:58:28 2012 CET
%\subtitle{There are 184 co-authors in this list}
\author{\small
Planck Collaboration:
N.~Aghanim\inst{54}
\and
M.~Arnaud\inst{69}\thanks{Corresponding author: M. Arnaud,  monique.arnaud@cea.fr}
\and
M.~Ashdown\inst{66, 5}
\and
F.~Atrio-Barandela\inst{17}
\and
J.~Aumont\inst{54}
\and
C.~Baccigalupi\inst{79}
\and
A.~Balbi\inst{34}
\and
A.~J.~Banday\inst{87, 8}
\and
R.~B.~Barreiro\inst{62}
% ...
}
\institute{\small
APC, AstroParticule et Cosmologie, Universit\'{e} Paris Diderot, CNRS/IN2P3, CEA/lrfu, Observatoire de Paris, Sorbonne Paris Cit\'{e}, 10, rue Alice Domon et L\'{e}onie Duquet, 75205 Paris Cedex 13, France\\
\and
Aalto University Mets\"{a}hovi Radio Observatory, Mets\"{a}hovintie 114, FIN-02540 Kylm\"{a}l\"{a}, Finland\\
\and
Agenzia Spaziale Italiana Science Data Center, c/o ESRIN, via Galileo Galilei, Frascati, Italy\\
\and
Argelander-Institut f\"{u}r Astronomie, Auf dem H\"{u}gel 71, D-53121 Bonn, Germany\\
\and
Astrophysics Group, Cavendish Laboratory, University of Cambridge, J J Thomson Avenue, Cambridge CB3 0HE, U.K.\\
\and
Atacama Large Millimeter/submillimeter Array, ALMA Santiago Central Offices, Alonso de Cordova 3107, Vitacura, Casilla 763 0355, Santiago, Chile\\
\and
CITA, University of Toronto, 60 St. George St., Toronto, ON M5S 3H8, Canada\\
\and
CNRS, IRAP, 9 Av. colonel Roche, BP 44346, F-31028 Toulouse cedex 4, France\\
\and
California Institute of Technology, Pasadena, California, U.S.A.\\
\and
Centre of Mathematics for Applications, University of Oslo, Blindern, Oslo, Norway\\
\and
Centro de Astrof\'{\i}sica, Universidade do Porto, Rua das Estrelas, 4150-762 Porto, Portugal\\
\and
Centro de Estudios de F\'{i}sica del Cosmos de Arag\'{o}n (CEFCA), Plaza San Juan, 1, planta 2, E-44001, Teruel, Spain\\
\and
Computational Cosmology Center, Lawrence Berkeley National Laboratory, Berkeley, California, U.S.A.\\
\and
Consejo Superior de Investigaciones Cient\'{\i}ficas (CSIC), Madrid, Spain\\
\and
DSM/Irfu/SPP, CEA-Saclay, F-91191 Gif-sur-Yvette Cedex, France\\
\and
DTU Space, National Space Institute, Juliane Mariesvej 30, Copenhagen, Denmark\\
\and
Departamento de F\'{\i}sica Fundamental, Facultad de Ciencias, Universidad de Salamanca, 37008 Salamanca, Spain\\
\and
Departamento de F\'{\i}sica, Universidad de Oviedo, Avda. Calvo Sotelo s/n, Oviedo, Spain\\
\and
Department of Astronomy and Astrophysics, University of Toronto, 50 Saint George Street, Toronto, Ontario, Canada\\
\and
Department of Physics \& Astronomy, University of British Columbia, 6224 Agricultural Road, Vancouver, British Columbia, Canada\\
\and
Department of Physics and Astronomy, University of Southern California, Los Angeles, California, U.S.A.\\
\and
Department of Physics and Astronomy, University of Sussex, Brighton BN1 9QH, U.K.\\
\and
Department of Physics, Gustaf H\"{a}llstr\"{o}min katu 2a, University of Helsinki, Helsinki, Finland\\
\and
Department of Physics, Purdue University, 525 Northwestern Avenue, West Lafayette, Indiana, U.S.A.\\
\and
Department of Physics, University of California, Berkeley, California, U.S.A.\\
\and
Department of Physics, University of California, One Shields Avenue, Davis, California, U.S.A.\\
\and
Department of Physics, University of California, Santa Barbara, California, U.S.A.\\
\and
Department of Physics, University of Illinois at Urbana-Champaign, 1110 West Green Street, Urbana, Illinois, U.S.A.\\
\and
Dipartimento di Fisica e Astronomia G. Galilei, Universit\`{a} degli Studi di Padova, via Marzolo 8, 35131 Padova, Italy\\
\and
Dipartimento di Fisica, Universit\`{a} La Sapienza, P. le A. Moro 2, Roma, Italy\\
\and
Dipartimento di Fisica, Universit\`{a} degli Studi di Milano, Via Celoria, 16, Milano, Italy\\
\and
Dipartimento di Fisica, Universit\`{a} degli Studi di Trieste, via A. Valerio 2, Trieste, Italy\\
\and
Dipartimento di Fisica, Universit\`{a} di Ferrara, Via Saragat 1, 44122 Ferrara, Italy\\
\and
Dipartimento di Fisica, Universit\`{a} di Roma Tor Vergata, Via della Ricerca Scientifica, 1, Roma, Italy\\
\and
Dipartimento di Matematica, Universit\`{a} di Roma Tor Vergata, Via della Ricerca Scientifica, 1, Roma, Italy\\
\and
Discovery Center, Niels Bohr Institute, Blegdamsvej 17, Copenhagen, Denmark\\
\and
Dpto. Astrof\'{i}sica, Universidad de La Laguna (ULL), E-38206 La Laguna, Tenerife, Spain\\
\and
European Southern Observatory, ESO Vitacura, Alonso de Cordova 3107, Vitacura, Casilla 19001, Santiago, Chile\\
\and
European Space Agency, ESAC, Camino bajo del Castillo, s/n, Urbanizaci\'{o}n Villafranca del Castillo, Villanueva de la Ca\~{n}ada, Madrid, Spain\\
\and
European Space Agency, ESAC, Planck Science Office, Camino bajo del Castillo, s/n, Urbanizaci\'{o}n Villafranca del Castillo, Villanueva de la Ca\~{n}ada, Madrid, Spain\\
\and
European Space Agency, ESTEC, Keplerlaan 1, 2201 AZ Noordwijk, The Netherlands\\
\and
Helsinki Institute of Physics, Gustaf H\"{a}llstr\"{o}min katu 2, University of Helsinki, Helsinki, Finland\\
\and
INAF - Osservatorio Astronomico di Padova, Vicolo dell'Osservatorio 5, Padova, Italy\\
\and
INAF - Osservatorio Astronomico di Roma, via di Frascati 33, Monte Porzio Catone, Italy\\
\and
INAF - Osservatorio Astronomico di Trieste, Via G.B. Tiepolo 11, Trieste, Italy\\
\and
INAF Istituto di Radioastronomia, Via P. Gobetti 101, 40129 Bologna, Italy\\
\and
INAF/IASF Bologna, Via Gobetti 101, Bologna, Italy\\
\and
INAF/IASF Milano, Via E. Bassini 15, Milano, Italy\\
\and
INFN, Sezione di Roma 1, Universit`{a} di Roma Sapienza, Piazzale Aldo Moro 2, 00185, Roma, Italy\\
\and
INRIA, Laboratoire de Recherche en Informatique, Universit\'{e} Paris-Sud 11, B\^{a}timent 490, 91405 Orsay Cedex, France\\
\and
IPAG: Institut de Plan\'{e}tologie et d'Astrophysique de Grenoble, Universit\'{e} Joseph Fourier, Grenoble 1 / CNRS-INSU, UMR 5274, Grenoble, F-38041, France\\
\and
Imperial College London, Astrophysics group, Blackett Laboratory, Prince Consort Road, London, SW7 2AZ, U.K.\\
\and
Infrared Processing and Analysis Center, California Institute of Technology, Pasadena, CA 91125, U.S.A.\\
\and
Institut d'Astrophysique Spatiale, CNRS (UMR8617) Universit\'{e} Paris-Sud 11, B\^{a}timent 121, Orsay, France\\
\and
Institut d'Astrophysique de Paris, CNRS (UMR7095), 98 bis Boulevard Arago, F-75014, Paris, France\\
\and
Institut de Ci\`{e}ncies de l'Espai, CSIC/IEEC, Facultat de Ci\`{e}ncies, Campus UAB, Torre C5 par-2, Bellaterra 08193, Spain\\
\and
Institute for Space Sciences, Bucharest-Magurale, Romania\\
\and
Institute of Astronomy and Astrophysics, Academia Sinica, Taipei, Taiwan\\
\and
Institute of Astronomy, University of Cambridge, Madingley Road, Cambridge CB3 0HA, U.K.\\
\and
Institute of Theoretical Astrophysics, University of Oslo, Blindern, Oslo, Norway\\
\and
Instituto de Astrof\'{\i}sica de Canarias, C/V\'{\i}a L\'{a}ctea s/n, La Laguna, Tenerife, Spain\\
\and
Instituto de F\'{\i}sica de Cantabria (CSIC-Universidad de Cantabria), Avda. de los Castros s/n, Santander, Spain\\
\and
Istituto di Fisica del Plasma, CNR-ENEA-EURATOM Association, Via R. Cozzi 53, Milano, Italy\\
\and
Jet Propulsion Laboratory, California Institute of Technology, 4800 Oak Grove Drive, Pasadena, California, U.S.A.\\
\and
Jodrell Bank Centre for Astrophysics, Alan Turing Building, School of Physics and Astronomy, The University of Manchester, Oxford Road, Manchester, M13 9PL, U.K.\\
\and
Kavli Institute for Cosmology Cambridge, Madingley Road, Cambridge, CB3 0HA, U.K.\\
\and
LAL, Universit\'{e} Paris-Sud, CNRS/IN2P3, Orsay, France\\
\and
LERMA, CNRS, Observatoire de Paris, 61 Avenue de l'Observatoire, Paris, France\\
\and
Laboratoire AIM, IRFU/Service d'Astrophysique - CEA/DSM - CNRS - Universit\'{e} Paris Diderot, B\^{a}t. 709, CEA-Saclay, F-91191 Gif-sur-Yvette Cedex, France\\
\and
Laboratoire Traitement et Communication de l'Information, CNRS (UMR 5141) and T\'{e}l\'{e}com ParisTech, 46 rue Barrault F-75634 Paris Cedex 13, France\\
\and
Laboratoire de Physique Subatomique et de Cosmologie, Universit\'{e} Joseph Fourier Grenoble I, CNRS/IN2P3, Institut National Polytechnique de Grenoble, 53 rue des Martyrs, 38026 Grenoble cedex, France\\
\and
Laboratoire de Physique Th\'{e}orique, Universit\'{e} Paris-Sud 11 \& CNRS, B\^{a}timent 210, 91405 Orsay, France\\
\and
Lawrence Berkeley National Laboratory, Berkeley, California, U.S.A.\\
\and
Max-Planck-Institut f\"{u}r Astrophysik, Karl-Schwarzschild-Str. 1, 85741 Garching, Germany\\
\and
Max-Planck-Institut f\"{u}r Extraterrestrische Physik, Giessenbachstra{\ss}e, 85748 Garching, Germany\\
\and
Niels Bohr Institute, Blegdamsvej 17, Copenhagen, Denmark\\
\and
Observational Cosmology, Mail Stop 367-17, California Institute of Technology, Pasadena, CA, 91125, U.S.A.\\
\and
Optical Science Laboratory, University College London, Gower Street, London, U.K.\\
\and
SISSA, Astrophysics Sector, via Bonomea 265, 34136, Trieste, Italy\\
\and
SUPA, Institute for Astronomy, University of Edinburgh, Royal Observatory, Blackford Hill, Edinburgh EH9 3HJ, U.K.\\
\and
School of Physics and Astronomy, Cardiff University, Queens Buildings, The Parade, Cardiff, CF24 3AA, U.K.\\
\and
Space Research Institute (IKI), Russian Academy of Sciences, Profsoyuznaya Str, 84/32, Moscow, 117997, Russia\\
\and
Space Sciences Laboratory, University of California, Berkeley, California, U.S.A.\\
\and
Stanford University, Dept of Physics, Varian Physics Bldg, 382 Via Pueblo Mall, Stanford, California, U.S.A.\\
\and
UPMC Univ Paris 06, UMR7095, 98 bis Boulevard Arago, F-75014, Paris, France\\
\and
Universit\"{a}t Heidelberg, Institut f\"{u}r Theoretische Astrophysik, Albert-\"{U}berle-Str. 2, 69120, Heidelberg, Germany\\
\and
Universit\'{e} de Toulouse, UPS-OMP, IRAP, F-31028 Toulouse cedex 4, France\\
% ...
}

%% file: Planck.tex
\def\setsymbol#1#2{\expandafter\def\csname #1\endcsname{#2}}
\def\getsymbol#1{\csname #1\endcsname}

%-----------------------------------------------------------------------
% Planck
%-----------------------------------------------------------------------
\def\Planck{{\it Planck\/}}

%-----------------------------------------------------------------------
% The Planck Helium-4 JT cooler
%-----------------------------------------------------------------------
\def\HeJT{$^4$He-JT}

%-----------------------------------------------------------------------
% To include all Planck Early Results papers in the reference lists
%-----------------------------------------------------------------------
\def\allearlypapers{\nocite{planck2011-1.1, planck2011-1.3, planck2011-1.4, planck2011-1.5, planck2011-1.6, planck2011-1.7, planck2011-1.10, planck2011-1.10sup, planck2011-5.1a, planck2011-5.1b, planck2011-5.2a, planck2011-5.2b, planck2011-5.2c, planck2011-6.1, planck2011-6.2, planck2011-6.3a, planck2011-6.4a, planck2011-6.4b, planck2011-6.6, planck2011-7.0, planck2011-7.2, planck2011-7.3, planck2011-7.7a, planck2011-7.7b, planck2011-7.12, planck2011-7.13}}

%-----------------------------------------------------------------------
% Tables
%-----------------------------------------------------------------------
\newbox\tablebox    \newdimen\tablewidth
\def\leaderfil{\leaders\hbox to 5pt{\hss.\hss}\hfil}
%
% use the following definition of \endPlancktable for ApJ style notes to tables, set to the %         width of the table
%\def\endPlancktable{\tablewidth=\wd\tablebox 
%
% use the following definition of \endPlancktable instead for A&A style notes set to full 
%         column width
\def\endPlancktable{\tablewidth=\columnwidth 
    $$\hss\copy\tablebox\hss$$
    \vskip-\lastskip\vskip -2pt}
\def\endPlancktablewide{\tablewidth=\textwidth 
    $$\hss\copy\tablebox\hss$$
    \vskip-\lastskip\vskip -2pt}
\def\tablenote#1 #2\par{\begingroup \parindent=0.8em
    \abovedisplayshortskip=0pt\belowdisplayshortskip=0pt
    \noindent
    $$\hss\vbox{\hsize\tablewidth \hangindent=\parindent \hangafter=1 \noindent
    \hbox to \parindent{\sup{\rm #1}\hss}\strut#2\strut\par}\hss$$
    \endgroup}
\def\doubleline{\vskip 3pt\hrule \vskip 1.5pt \hrule \vskip 5pt}

%-----------------------------------------------------------------------
% useful macros
%-----------------------------------------------------------------------
%
\def\L2{\ifmmode L_2\else $L_2$\fi}
\def\dtt{\Delta T/T}
\def\DeltaT{\ifmmode \Delta T\else $\Delta T$\fi}
\def\deltat{\ifmmode \Delta t\else $\Delta t$\fi}
\def\fknee{\ifmmode f_{\rm knee}\else $f_{\rm knee}$\fi}
\def\Fmax{\ifmmode F_{\rm max}\else $F_{\rm max}$\fi}
\def\solar{\ifmmode{\rm M}_{\mathord\odot}\else${\rm M}_{\mathord\odot}$\fi}
\def\sol{\solar}
\def\mag{\sup{m}}
\def\inv{\ifmmode^{-1}\else$^{-1}$\fi}
\def\mo{\ifmmode^{-1}\else$^{-1}$\fi}
\def\sup#1{\ifmmode ^{\rm #1}\else $^{\rm #1}$\fi}
\def\expo#1{\ifmmode \times 10^{#1}\else $\times 10^{#1}$\fi}
\def\,{\thinspace}
\def\lsim{\mathrel{\raise .4ex\hbox{\rlap{$<$}\lower 1.2ex\hbox{$\sim$}}}}
\def\gsim{\mathrel{\raise .4ex\hbox{\rlap{$>$}\lower 1.2ex\hbox{$\sim$}}}}
\let\lea=\lsim
\let\gea=\gsim
\def\simprop{\mathrel{\raise .4ex\hbox{\rlap{$\propto$}\lower 1.2ex\hbox{$\sim$}}}}
\def\deg{\ifmmode^\circ\else$^\circ$\fi}
\def\pdeg{\ifmmode $\setbox0=\hbox{$^{\circ}$}\rlap{\hskip.11\wd0 .}$^{\circ}
          \else \setbox0=\hbox{$^{\circ}$}\rlap{\hskip.11\wd0 .}$^{\circ}$\fi}
\def\arcs{\ifmmode {^{\scriptstyle\prime\prime}}
          \else $^{\scriptstyle\prime\prime}$\fi}
\def\arcm{\ifmmode {^{\scriptstyle\prime}}
          \else $^{\scriptstyle\prime}$\fi}
\newdimen\sa  \newdimen\sb
\def\parcs{\sa=.07em \sb=.03em
     \ifmmode \hbox{\rlap{.}}^{\scriptstyle\prime\kern -\sb\prime}\hbox{\kern -\sa}
     \else \rlap{.}$^{\scriptstyle\prime\kern -\sb\prime}$\kern -\sa\fi}
\def\parcm{\sa=.08em \sb=.03em
     \ifmmode \hbox{\rlap{.}\kern\sa}^{\scriptstyle\prime}\hbox{\kern-\sb}
     \else \rlap{.}\kern\sa$^{\scriptstyle\prime}$\kern-\sb\fi}
\def\ra[#1 #2 #3.#4]{#1\sup{h}#2\sup{m}#3\sup{s}\llap.#4}
\def\dec[#1 #2 #3.#4]{#1\deg#2\arcm#3\arcs\llap.#4}
\def\deco[#1 #2 #3]{#1\deg#2\arcm#3\arcs}
\def\rra[#1 #2]{#1\sup{h}#2\sup{m}}
\def\page{\vfill\eject}
\def\dots{\relax\ifmmode \ldots\else $\ldots$\fi}
%
%-----------------------------------------------------------------------
% units
%-----------------------------------------------------------------------
%
\def\WHzsr{\ifmmode $W\,Hz\mo\,sr\mo$\else W\,Hz\mo\,sr\mo\fi}
\def\mHz{\ifmmode $\,mHz$\else \,mHz\fi}
\def\GHz{\ifmmode $\,GHz$\else \,GHz\fi}
\def\mKs{\ifmmode $\,mK\,s$^{1/2}\else \,mK\,s$^{1/2}$\fi}
\def\muKs{\ifmmode \,\mu$K\,s$^{1/2}\else \,$\mu$K\,s$^{1/2}$\fi}
\def\muKRJs{\ifmmode \,\mu$K$_{\rm RJ}$\,s$^{1/2}\else \,$\mu$K$_{\rm RJ}$\,s$^{1/2}$\fi}
\def\muKHz{\ifmmode \,\mu$K\,Hz$^{-1/2}\else \,$\mu$K\,Hz$^{-1/2}$\fi}
\def\MJysr{\ifmmode \,$MJy\,sr\mo$\else \,MJy\,sr\mo\fi}
\def\MJysrmK{\ifmmode \,$MJy\,sr\mo$\,mK$_{\rm CMB}\mo\else \,MJy\,sr\mo\,mK$_{\rm CMB}\mo$\fi}
\def\microns{\ifmmode \,\mu$m$\else \,$\mu$m\fi}
\def\micron{\microns}
\def\muK{\ifmmode \,\mu$K$\else \,$\mu$\hbox{K}\fi}
\def\microK{\ifmmode \,\mu$K$\else \,$\mu$\hbox{K}\fi}
\def\muW{\ifmmode \,\mu$W$\else \,$\mu$\hbox{W}\fi}
\def\kms{\ifmmode $\,km\,s$^{-1}\else \,km\,s$^{-1}$\fi}
\def\kmsMpc{\ifmmode $\,\kms\,Mpc\mo$\else \,\kms\,Mpc\mo\fi}
%
%
%----------------------------------------------------------------------

% LFI Center Frequency

\setsymbol{LFI:center:frequency:70GHz:units}{70.3\,GHz}
\setsymbol{LFI:center:frequency:44GHz:units}{44.1\,GHz}
\setsymbol{LFI:center:frequency:30GHz:units}{28.5\,GHz}

\setsymbol{LFI:center:frequency:70GHz}{70.3}
\setsymbol{LFI:center:frequency:44GHz}{44.1}
\setsymbol{LFI:center:frequency:30GHz}{28.5}

\setsymbol{LFI:center:frequency:LFI18:Rad:M:units}{71.7\GHz}
\setsymbol{LFI:center:frequency:LFI19:Rad:M:units}{67.5\GHz}
\setsymbol{LFI:center:frequency:LFI20:Rad:M:units}{69.2\GHz}
\setsymbol{LFI:center:frequency:LFI21:Rad:M:units}{70.4\GHz}
\setsymbol{LFI:center:frequency:LFI22:Rad:M:units}{71.5\GHz}
\setsymbol{LFI:center:frequency:LFI23:Rad:M:units}{70.8\GHz}
\setsymbol{LFI:center:frequency:LFI24:Rad:M:units}{44.4\GHz}
\setsymbol{LFI:center:frequency:LFI25:Rad:M:units}{44.0\GHz}
\setsymbol{LFI:center:frequency:LFI26:Rad:M:units}{43.9\GHz}
\setsymbol{LFI:center:frequency:LFI27:Rad:M:units}{28.3\GHz}
\setsymbol{LFI:center:frequency:LFI28:Rad:M:units}{28.8\GHz}
\setsymbol{LFI:center:frequency:LFI18:Rad:S:units}{70.1\GHz}
\setsymbol{LFI:center:frequency:LFI19:Rad:S:units}{69.6\GHz}
\setsymbol{LFI:center:frequency:LFI20:Rad:S:units}{69.5\GHz}
\setsymbol{LFI:center:frequency:LFI21:Rad:S:units}{69.5\GHz}
\setsymbol{LFI:center:frequency:LFI22:Rad:S:units}{72.8\GHz}
\setsymbol{LFI:center:frequency:LFI23:Rad:S:units}{71.3\GHz}
\setsymbol{LFI:center:frequency:LFI24:Rad:S:units}{44.1\GHz}
\setsymbol{LFI:center:frequency:LFI25:Rad:S:units}{44.1\GHz}
\setsymbol{LFI:center:frequency:LFI26:Rad:S:units}{44.1\GHz}
\setsymbol{LFI:center:frequency:LFI27:Rad:S:units}{28.5\GHz}
\setsymbol{LFI:center:frequency:LFI28:Rad:S:units}{28.2\GHz}

\setsymbol{LFI:center:frequency:LFI18:Rad:M}{71.7}
\setsymbol{LFI:center:frequency:LFI19:Rad:M}{67.5}
\setsymbol{LFI:center:frequency:LFI20:Rad:M}{69.2}
\setsymbol{LFI:center:frequency:LFI21:Rad:M}{70.4}
\setsymbol{LFI:center:frequency:LFI22:Rad:M}{71.5}
\setsymbol{LFI:center:frequency:LFI23:Rad:M}{70.8}
\setsymbol{LFI:center:frequency:LFI24:Rad:M}{44.4}
\setsymbol{LFI:center:frequency:LFI25:Rad:M}{44.0}
\setsymbol{LFI:center:frequency:LFI26:Rad:M}{43.9}
\setsymbol{LFI:center:frequency:LFI27:Rad:M}{28.3}
\setsymbol{LFI:center:frequency:LFI28:Rad:M}{28.8}
\setsymbol{LFI:center:frequency:LFI18:Rad:S}{70.1}
\setsymbol{LFI:center:frequency:LFI19:Rad:S}{69.6}
\setsymbol{LFI:center:frequency:LFI20:Rad:S}{69.5}
\setsymbol{LFI:center:frequency:LFI21:Rad:S}{69.5}
\setsymbol{LFI:center:frequency:LFI22:Rad:S}{72.8}
\setsymbol{LFI:center:frequency:LFI23:Rad:S}{71.3}
\setsymbol{LFI:center:frequency:LFI24:Rad:S}{44.1}
\setsymbol{LFI:center:frequency:LFI25:Rad:S}{44.1}
\setsymbol{LFI:center:frequency:LFI26:Rad:S}{44.1}
\setsymbol{LFI:center:frequency:LFI27:Rad:S}{28.5}
\setsymbol{LFI:center:frequency:LFI28:Rad:S}{28.2}

% LFI White Noise Sensitivity

\setsymbol{LFI:white:noise:sensitivity:70GHz:units}{152.6\muKs}
\setsymbol{LFI:white:noise:sensitivity:44GHz:units}{173.1\muKs}
\setsymbol{LFI:white:noise:sensitivity:30GHz:units}{146.8\muKs}

\setsymbol{LFI:white:noise:sensitivity:70GHz}{152.6}
\setsymbol{LFI:white:noise:sensitivity:44GHz}{173.1}
\setsymbol{LFI:white:noise:sensitivity:30GHz}{146.8}

\setsymbol{LFI:white:noise:sensitivity:LFI18:Rad:M:units}{512.0\muKs}
\setsymbol{LFI:white:noise:sensitivity:LFI19:Rad:M:units}{581.4\muKs}
\setsymbol{LFI:white:noise:sensitivity:LFI20:Rad:M:units}{590.8\muKs}
\setsymbol{LFI:white:noise:sensitivity:LFI21:Rad:M:units}{455.2\muKs}
\setsymbol{LFI:white:noise:sensitivity:LFI22:Rad:M:units}{492.0\muKs}
\setsymbol{LFI:white:noise:sensitivity:LFI23:Rad:M:units}{507.7\muKs}
\setsymbol{LFI:white:noise:sensitivity:LFI24:Rad:M:units}{462.2\muKs}
\setsymbol{LFI:white:noise:sensitivity:LFI25:Rad:M:units}{413.6\muKs}
\setsymbol{LFI:white:noise:sensitivity:LFI26:Rad:M:units}{478.6\muKs}
\setsymbol{LFI:white:noise:sensitivity:LFI27:Rad:M:units}{277.7\muKs}
\setsymbol{LFI:white:noise:sensitivity:LFI28:Rad:M:units}{312.3\muKs}
\setsymbol{LFI:white:noise:sensitivity:LFI18:Rad:S:units}{465.7\muKs}
\setsymbol{LFI:white:noise:sensitivity:LFI19:Rad:S:units}{555.6\muKs}
\setsymbol{LFI:white:noise:sensitivity:LFI20:Rad:S:units}{623.2\muKs}
\setsymbol{LFI:white:noise:sensitivity:LFI21:Rad:S:units}{564.1\muKs}
\setsymbol{LFI:white:noise:sensitivity:LFI22:Rad:S:units}{534.4\muKs}
\setsymbol{LFI:white:noise:sensitivity:LFI23:Rad:S:units}{542.4\muKs}
\setsymbol{LFI:white:noise:sensitivity:LFI24:Rad:S:units}{399.2\muKs}
\setsymbol{LFI:white:noise:sensitivity:LFI25:Rad:S:units}{392.6\muKs}
\setsymbol{LFI:white:noise:sensitivity:LFI26:Rad:S:units}{418.6\muKs}
\setsymbol{LFI:white:noise:sensitivity:LFI27:Rad:S:units}{302.9\muKs}
\setsymbol{LFI:white:noise:sensitivity:LFI28:Rad:S:units}{285.3\muKs}

\setsymbol{LFI:white:noise:sensitivity:LFI18:Rad:M}{512.0}
\setsymbol{LFI:white:noise:sensitivity:LFI19:Rad:M}{581.4}
\setsymbol{LFI:white:noise:sensitivity:LFI20:Rad:M}{590.8}
\setsymbol{LFI:white:noise:sensitivity:LFI21:Rad:M}{455.2}
\setsymbol{LFI:white:noise:sensitivity:LFI22:Rad:M}{492.0}
\setsymbol{LFI:white:noise:sensitivity:LFI23:Rad:M}{507.7}
\setsymbol{LFI:white:noise:sensitivity:LFI24:Rad:M}{462.2}
\setsymbol{LFI:white:noise:sensitivity:LFI25:Rad:M}{413.6}
\setsymbol{LFI:white:noise:sensitivity:LFI26:Rad:M}{478.6}
\setsymbol{LFI:white:noise:sensitivity:LFI27:Rad:M}{277.7}
\setsymbol{LFI:white:noise:sensitivity:LFI28:Rad:M}{312.3}
\setsymbol{LFI:white:noise:sensitivity:LFI18:Rad:S}{465.7}
\setsymbol{LFI:white:noise:sensitivity:LFI19:Rad:S}{555.6}
\setsymbol{LFI:white:noise:sensitivity:LFI20:Rad:S}{623.2}
\setsymbol{LFI:white:noise:sensitivity:LFI21:Rad:S}{564.1}
\setsymbol{LFI:white:noise:sensitivity:LFI22:Rad:S}{534.4}
\setsymbol{LFI:white:noise:sensitivity:LFI23:Rad:S}{542.4}
\setsymbol{LFI:white:noise:sensitivity:LFI24:Rad:S}{399.2}
\setsymbol{LFI:white:noise:sensitivity:LFI25:Rad:S}{392.6}
\setsymbol{LFI:white:noise:sensitivity:LFI26:Rad:S}{418.6}
\setsymbol{LFI:white:noise:sensitivity:LFI27:Rad:S}{302.9}
\setsymbol{LFI:white:noise:sensitivity:LFI28:Rad:S}{285.3}

% LFI Knee Frequency

\setsymbol{LFI:knee:frequency:70GHz:units}{29.5\mHz}
\setsymbol{LFI:knee:frequency:44GHz:units}{56.2\mHz}
\setsymbol{LFI:knee:frequency:30GHz:units}{113.7\mHz}

\setsymbol{LFI:knee:frequency:70GHz}{29.5}
\setsymbol{LFI:knee:frequency:44GHz}{56.2}
\setsymbol{LFI:knee:frequency:30GHz}{113.7}

\setsymbol{LFI:knee:frequency:LFI18:Rad:M:units}{16.3\mHz}
\setsymbol{LFI:knee:frequency:LFI19:Rad:M:units}{15.1\mHz}
\setsymbol{LFI:knee:frequency:LFI20:Rad:M:units}{18.7\mHz}
\setsymbol{LFI:knee:frequency:LFI21:Rad:M:units}{37.2\mHz}
\setsymbol{LFI:knee:frequency:LFI22:Rad:M:units}{12.7\mHz}
\setsymbol{LFI:knee:frequency:LFI23:Rad:M:units}{34.6\mHz}
\setsymbol{LFI:knee:frequency:LFI24:Rad:M:units}{46.2\mHz}
\setsymbol{LFI:knee:frequency:LFI25:Rad:M:units}{24.9\mHz}
\setsymbol{LFI:knee:frequency:LFI26:Rad:M:units}{67.6\mHz}
\setsymbol{LFI:knee:frequency:LFI27:Rad:M:units}{187.4\mHz}
\setsymbol{LFI:knee:frequency:LFI28:Rad:M:units}{122.2\mHz}
\setsymbol{LFI:knee:frequency:LFI18:Rad:S:units}{17.7\mHz}
\setsymbol{LFI:knee:frequency:LFI19:Rad:S:units}{22.0\mHz}
\setsymbol{LFI:knee:frequency:LFI20:Rad:S:units}{8.7\mHz}
\setsymbol{LFI:knee:frequency:LFI21:Rad:S:units}{25.9\mHz}
\setsymbol{LFI:knee:frequency:LFI22:Rad:S:units}{15.8\mHz}
\setsymbol{LFI:knee:frequency:LFI23:Rad:S:units}{129.8\mHz}
\setsymbol{LFI:knee:frequency:LFI24:Rad:S:units}{100.9\mHz}
\setsymbol{LFI:knee:frequency:LFI25:Rad:S:units}{38.9\mHz}
\setsymbol{LFI:knee:frequency:LFI26:Rad:S:units}{58.9\mHz}
\setsymbol{LFI:knee:frequency:LFI27:Rad:S:units}{104.4\mHz}
\setsymbol{LFI:knee:frequency:LFI28:Rad:S:units}{40.7\mHz}

\setsymbol{LFI:knee:frequency:LFI18:Rad:M}{16.3}
\setsymbol{LFI:knee:frequency:LFI19:Rad:M}{15.1}
\setsymbol{LFI:knee:frequency:LFI20:Rad:M}{18.7}
\setsymbol{LFI:knee:frequency:LFI21:Rad:M}{37.2}
\setsymbol{LFI:knee:frequency:LFI22:Rad:M}{12.7}
\setsymbol{LFI:knee:frequency:LFI23:Rad:M}{34.6}
\setsymbol{LFI:knee:frequency:LFI24:Rad:M}{46.2}
\setsymbol{LFI:knee:frequency:LFI25:Rad:M}{24.9}
\setsymbol{LFI:knee:frequency:LFI26:Rad:M}{67.6}
\setsymbol{LFI:knee:frequency:LFI27:Rad:M}{187.4}
\setsymbol{LFI:knee:frequency:LFI28:Rad:M}{122.2}
\setsymbol{LFI:knee:frequency:LFI18:Rad:S}{17.7}
\setsymbol{LFI:knee:frequency:LFI19:Rad:S}{22.0}
\setsymbol{LFI:knee:frequency:LFI20:Rad:S}{8.7}
\setsymbol{LFI:knee:frequency:LFI21:Rad:S}{25.9}
\setsymbol{LFI:knee:frequency:LFI22:Rad:S}{15.8}
\setsymbol{LFI:knee:frequency:LFI23:Rad:S}{129.8}
\setsymbol{LFI:knee:frequency:LFI24:Rad:S}{100.9}
\setsymbol{LFI:knee:frequency:LFI25:Rad:S}{38.9}
\setsymbol{LFI:knee:frequency:LFI26:Rad:S}{58.9}
\setsymbol{LFI:knee:frequency:LFI27:Rad:S}{104.4}
\setsymbol{LFI:knee:frequency:LFI28:Rad:S}{40.7}

% LFI low frequency noise slope

\setsymbol{LFI:slope:70GHz:units}{$-1.03$\mHz}
\setsymbol{LFI:slope:44GHz:units}{$-0.89$\mHz}
\setsymbol{LFI:slope:30GHz:units}{$-0.87$\mHz}

\setsymbol{LFI:slope:70GHz}{$-1.03$}
\setsymbol{LFI:slope:44GHz}{$-0.89$}
\setsymbol{LFI:slope:30GHz}{$-0.87$}

\setsymbol{LFI:slope:LFI18:Rad:M:units}{$-1.04$\mHz}
\setsymbol{LFI:slope:LFI19:Rad:M:units}{$-1.09$\mHz}
\setsymbol{LFI:slope:LFI20:Rad:M:units}{$-0.69$\mHz}
\setsymbol{LFI:slope:LFI21:Rad:M:units}{$-1.56$\mHz}
\setsymbol{LFI:slope:LFI22:Rad:M:units}{$-1.01$\mHz}
\setsymbol{LFI:slope:LFI23:Rad:M:units}{$-0.96$\mHz}
\setsymbol{LFI:slope:LFI24:Rad:M:units}{$-0.83$\mHz}
\setsymbol{LFI:slope:LFI25:Rad:M:units}{$-0.91$\mHz}
\setsymbol{LFI:slope:LFI26:Rad:M:units}{$-0.95$\mHz}
\setsymbol{LFI:slope:LFI27:Rad:M:units}{$-0.87$\mHz}
\setsymbol{LFI:slope:LFI28:Rad:M:units}{$-0.88$\mHz}
\setsymbol{LFI:slope:LFI18:Rad:S:units}{$-1.15$\mHz}
\setsymbol{LFI:slope:LFI19:Rad:S:units}{$-1.00$\mHz}
\setsymbol{LFI:slope:LFI20:Rad:S:units}{$-0.95$\mHz}
\setsymbol{LFI:slope:LFI21:Rad:S:units}{$-0.92$\mHz}
\setsymbol{LFI:slope:LFI22:Rad:S:units}{$-1.01$\mHz}
\setsymbol{LFI:slope:LFI23:Rad:S:units}{$-0.95$\mHz}
\setsymbol{LFI:slope:LFI24:Rad:S:units}{$-0.73$\mHz}
\setsymbol{LFI:slope:LFI25:Rad:S:units}{$-1.16$\mHz}
\setsymbol{LFI:slope:LFI26:Rad:S:units}{$-0.79$\mHz}
\setsymbol{LFI:slope:LFI27:Rad:S:units}{$-0.82$\mHz}
\setsymbol{LFI:slope:LFI28:Rad:S:units}{$-0.91$\mHz}

\setsymbol{LFI:slope:LFI18:Rad:M}{$-1.04$}
\setsymbol{LFI:slope:LFI19:Rad:M}{$-1.09$}
\setsymbol{LFI:slope:LFI20:Rad:M}{$-0.69$}
\setsymbol{LFI:slope:LFI21:Rad:M}{$-1.56$}
\setsymbol{LFI:slope:LFI22:Rad:M}{$-1.01$}
\setsymbol{LFI:slope:LFI23:Rad:M}{$-0.96$}
\setsymbol{LFI:slope:LFI24:Rad:M}{$-0.83$}
\setsymbol{LFI:slope:LFI25:Rad:M}{$-0.91$}
\setsymbol{LFI:slope:LFI26:Rad:M}{$-0.95$}
\setsymbol{LFI:slope:LFI27:Rad:M}{$-0.87$}
\setsymbol{LFI:slope:LFI28:Rad:M}{$-0.88$}
\setsymbol{LFI:slope:LFI18:Rad:S}{$-1.15$}
\setsymbol{LFI:slope:LFI19:Rad:S}{$-1.00$}
\setsymbol{LFI:slope:LFI20:Rad:S}{$-0.95$}
\setsymbol{LFI:slope:LFI21:Rad:S}{$-0.92$}
\setsymbol{LFI:slope:LFI22:Rad:S}{$-1.01$}
\setsymbol{LFI:slope:LFI23:Rad:S}{$-0.95$}
\setsymbol{LFI:slope:LFI24:Rad:S}{$-0.73$}
\setsymbol{LFI:slope:LFI25:Rad:S}{$-1.16$}
\setsymbol{LFI:slope:LFI26:Rad:S}{$-0.79$}
\setsymbol{LFI:slope:LFI27:Rad:S}{$-0.82$}
\setsymbol{LFI:slope:LFI28:Rad:S}{$-0.91$}

% LFI Beam FWHM

\setsymbol{LFI:FWHM:70GHz:units}{13\parcm01}
\setsymbol{LFI:FWHM:44GHz:units}{27\parcm92}
\setsymbol{LFI:FWHM:30GHz:units}{32\parcm65}

\setsymbol{LFI:FWHM:70GHz}{13.01}
\setsymbol{LFI:FWHM:44GHz}{27.92}
\setsymbol{LFI:FWHM:30GHz}{32.65}

\setsymbol{LFI:FWHM:LFI18:units}{13\parcm39}
\setsymbol{LFI:FWHM:LFI19:units}{13\parcm01}
\setsymbol{LFI:FWHM:LFI20:units}{12\parcm75}
\setsymbol{LFI:FWHM:LFI21:units}{12\parcm74}
\setsymbol{LFI:FWHM:LFI22:units}{12\parcm87}
\setsymbol{LFI:FWHM:LFI23:units}{13\parcm27}
\setsymbol{LFI:FWHM:LFI24:units}{22\parcm98}
\setsymbol{LFI:FWHM:LFI25:units}{30\parcm46}
\setsymbol{LFI:FWHM:LFI26:units}{30\parcm31}
\setsymbol{LFI:FWHM:LFI27:units}{32\parcm65}
\setsymbol{LFI:FWHM:LFI28:units}{32\parcm66}

\setsymbol{LFI:FWHM:LFI18}{13.39}
\setsymbol{LFI:FWHM:LFI19}{13.01}
\setsymbol{LFI:FWHM:LFI20}{12.75}
\setsymbol{LFI:FWHM:LFI21}{12.74}
\setsymbol{LFI:FWHM:LFI22}{12.87}
\setsymbol{LFI:FWHM:LFI23}{13.27}
\setsymbol{LFI:FWHM:LFI24}{22.98}
\setsymbol{LFI:FWHM:LFI25}{30.46}
\setsymbol{LFI:FWHM:LFI26}{30.31}
\setsymbol{LFI:FWHM:LFI27}{32.65}
\setsymbol{LFI:FWHM:LFI28}{32.66}

% LFI Beam FWHM Uncertainty
% When uncertainties are routinely available for all quantities, we'll likely change the format to build them into 
% the \setsymbol command.  For now, this will be a bit easier.

%\setsymbol{LFI:FWHM:uncertainty:70GHz}{TBD\arcm}
%\setsymbol{LFI:FWHM:uncertainty:44GHz}{TBD\arcm}
%\setsymbol{LFI:FWHM:uncertainty:30GHz}{TBD\arcm}

\setsymbol{LFI:FWHM:uncertainty:LFI18:units}{0.170\arcm}
\setsymbol{LFI:FWHM:uncertainty:LFI19:units}{0.174\arcm}
\setsymbol{LFI:FWHM:uncertainty:LFI20:units}{0.170\arcm}
\setsymbol{LFI:FWHM:uncertainty:LFI21:units}{0.156\arcm}
\setsymbol{LFI:FWHM:uncertainty:LFI22:units}{0.164\arcm}
\setsymbol{LFI:FWHM:uncertainty:LFI23:units}{0.171\arcm}
\setsymbol{LFI:FWHM:uncertainty:LFI24:units}{0.652\arcm}
\setsymbol{LFI:FWHM:uncertainty:LFI25:units}{1.075\arcm}
\setsymbol{LFI:FWHM:uncertainty:LFI26:units}{1.131\arcm}
\setsymbol{LFI:FWHM:uncertainty:LFI27:units}{1.266\arcm}
\setsymbol{LFI:FWHM:uncertainty:LFI28:units}{1.287\arcm}

\setsymbol{LFI:FWHM:uncertainty:LFI18}{0.170}
\setsymbol{LFI:FWHM:uncertainty:LFI19}{0.174}
\setsymbol{LFI:FWHM:uncertainty:LFI20}{0.170}
\setsymbol{LFI:FWHM:uncertainty:LFI21}{0.156}
\setsymbol{LFI:FWHM:uncertainty:LFI22}{0.164}
\setsymbol{LFI:FWHM:uncertainty:LFI23}{0.171}
\setsymbol{LFI:FWHM:uncertainty:LFI24}{0.652}
\setsymbol{LFI:FWHM:uncertainty:LFI25}{1.075}
\setsymbol{LFI:FWHM:uncertainty:LFI26}{1.131}
\setsymbol{LFI:FWHM:uncertainty:LFI27}{1.266}
\setsymbol{LFI:FWHM:uncertainty:LFI28}{1.287}

% HFI Center Frequency

\setsymbol{HFI:center:frequency:100GHz:units}{100\,GHz}
\setsymbol{HFI:center:frequency:143GHz:units}{143\,GHz}
\setsymbol{HFI:center:frequency:217GHz:units}{217\,GHz}
\setsymbol{HFI:center:frequency:353GHz:units}{353\,GHz}
\setsymbol{HFI:center:frequency:545GHz:units}{545\,GHz}
\setsymbol{HFI:center:frequency:857GHz:units}{857\,GHz}

\setsymbol{HFI:center:frequency:100GHz}{100}
\setsymbol{HFI:center:frequency:143GHz}{143}
\setsymbol{HFI:center:frequency:217GHz}{217}
\setsymbol{HFI:center:frequency:353GHz}{353}
\setsymbol{HFI:center:frequency:545GHz}{545}
\setsymbol{HFI:center:frequency:857GHz}{857}

% HFI Number of Detectors

\setsymbol{HFI:Ndetectors:100GHz}{8}
\setsymbol{HFI:Ndetectors:143GHz}{11}
\setsymbol{HFI:Ndetectors:217GHz}{12}
\setsymbol{HFI:Ndetectors:353GHz}{12}
\setsymbol{HFI:Ndetectors:545GHz}{3}
\setsymbol{HFI:Ndetectors:857GHz}{4}

% HFI FWHM on maps

\setsymbol{HFI:FWHM:Maps:100GHz:units}{9\parcm88}
\setsymbol{HFI:FWHM:Maps:143GHz:units}{7\parcm18}
\setsymbol{HFI:FWHM:Maps:217GHz:units}{4\parcm87}
\setsymbol{HFI:FWHM:Maps:353GHz:units}{4\parcm65}
\setsymbol{HFI:FWHM:Maps:545GHz:units}{4\parcm72}
\setsymbol{HFI:FWHM:Maps:857GHz:units}{4\parcm39}
\setsymbol{HFI:FWHM:Maps:100GHz}{9.88}
\setsymbol{HFI:FWHM:Maps:143GHz}{7.18}
\setsymbol{HFI:FWHM:Maps:217GHz}{4.87}
\setsymbol{HFI:FWHM:Maps:353GHz}{4.65}
\setsymbol{HFI:FWHM:Maps:545GHz}{4.72}
\setsymbol{HFI:FWHM:Maps:857GHz}{4.39}

% HFI Beam Ellipticity on maps

\setsymbol{HFI:beam:ellipticity:Maps:100GHz}{1.15}
\setsymbol{HFI:beam:ellipticity:Maps:143GHz}{1.01}
\setsymbol{HFI:beam:ellipticity:Maps:217GHz}{1.06}
\setsymbol{HFI:beam:ellipticity:Maps:353GHz}{1.05}
\setsymbol{HFI:beam:ellipticity:Maps:545GHz}{1.14}
\setsymbol{HFI:beam:ellipticity:Maps:857GHz}{1.19}

% HFI optical Beam FWHM from Mars; time response deconvolved: frequency  average of values in table 4 in HFI instrument paper

\setsymbol{HFI:FWHM:Mars:100GHz:units}{9\parcm37}
\setsymbol{HFI:FWHM:Mars:143GHz:units}{7\parcm04}
\setsymbol{HFI:FWHM:Mars:217GHz:units}{4\parcm68}
\setsymbol{HFI:FWHM:Mars:353GHz:units}{4\parcm43}
\setsymbol{HFI:FWHM:Mars:545GHz:units}{3\parcm80}
\setsymbol{HFI:FWHM:Mars:857GHz:units}{3\parcm67}

\setsymbol{HFI:FWHM:Mars:100GHz}{9.37}
\setsymbol{HFI:FWHM:Mars:143GHz}{7.04}
\setsymbol{HFI:FWHM:Mars:217GHz}{4.68}
\setsymbol{HFI:FWHM:Mars:353GHz}{4.43}
\setsymbol{HFI:FWHM:Mars:545GHz}{3.80}
\setsymbol{HFI:FWHM:Mars:857GHz}{3.67}

% HFI optical Beam Ellipticity from Mars; time response deconvolved: frequency average of values in table 4 in HFI instrument paper

\setsymbol{HFI:beam:ellipticity:Mars:100GHz}{1.18}
\setsymbol{HFI:beam:ellipticity:Mars:143GHz}{1.03}
\setsymbol{HFI:beam:ellipticity:Mars:217GHz}{1.14}
\setsymbol{HFI:beam:ellipticity:Mars:353GHz}{1.09}
\setsymbol{HFI:beam:ellipticity:Mars:545GHz}{1.25}
\setsymbol{HFI:beam:ellipticity:Mars:857GHz}{1.03}

% HFI CMB relative calibration accuracy

\setsymbol{HFI:CMB:relative:calibration:100GHz}{$\lsim 1\%$}
\setsymbol{HFI:CMB:relative:calibration:143GHz}{$\lsim 1\%$}
\setsymbol{HFI:CMB:relative:calibration:217GHz}{$\lsim 1\%$}
\setsymbol{HFI:CMB:relative:calibration:353GHz}{$\lsim 1\%$}
\setsymbol{HFI:CMB:relative:calibration:545GHz}{}
\setsymbol{HFI:CMB:relative:calibration:857GHz}{}

% HFI CMB absolute calibration accuracy

\setsymbol{HFI:CMB:absolute:calibration:100GHz}{$\lsim 2\%$}
\setsymbol{HFI:CMB:absolute:calibration:143GHz}{$\lsim 2\%$}
\setsymbol{HFI:CMB:absolute:calibration:217GHz}{$\lsim 2\%$}
\setsymbol{HFI:CMB:absolute:calibration:353GHz}{$\lsim 2\%$}
\setsymbol{HFI:CMB:absolute:calibration:545GHz}{}
\setsymbol{HFI:CMB:absolute:calibration:857GHz}{}

% HFI FIRAS gain calibration accuracy: statistical

\setsymbol{HFI:FIRAS:gain:calibration:accuracy:statistical:100GHz}{}
\setsymbol{HFI:FIRAS:gain:calibration:accuracy:statistical:143GHz}{}
\setsymbol{HFI:FIRAS:gain:calibration:accuracy:statistical:217GHz}{}
\setsymbol{HFI:FIRAS:gain:calibration:accuracy:statistical:353GHz}{2.5\%}
\setsymbol{HFI:FIRAS:gain:calibration:accuracy:statistical:545GHz}{1\%}
\setsymbol{HFI:FIRAS:gain:calibration:accuracy:statistical:857GHz}{0.5\%}

% HFI FIRAS gain calibration accuracy: systematic

\setsymbol{HFI:FIRAS:gain:calibration:accuracy:systematic:100GHz}{}
\setsymbol{HFI:FIRAS:gain:calibration:accuracy:systematic:143GHz}{}
\setsymbol{HFI:FIRAS:gain:calibration:accuracy:systematic:217GHz}{}
\setsymbol{HFI:FIRAS:gain:calibration:accuracy:systematic:353GHz}{}
\setsymbol{HFI:FIRAS:gain:calibration:accuracy:systematic:545GHz}{7\%}
\setsymbol{HFI:FIRAS:gain:calibration:accuracy:systematic:857GHz}{7\%}

% HFI FIRAS zero point accuracy:

\setsymbol{HFI:FIRAS:zero:point:accuracy:100GHz:units}{0.8\MJysr}
\setsymbol{HFI:FIRAS:zero:point:accuracy:143GHz:units}{}
\setsymbol{HFI:FIRAS:zero:point:accuracy:217GHz:units}{}
\setsymbol{HFI:FIRAS:zero:point:accuracy:353GHz:units}{1.4\MJysr}
\setsymbol{HFI:FIRAS:zero:point:accuracy:545GHz:units}{2.2\MJysr}
\setsymbol{HFI:FIRAS:zero:point:accuracy:857GHz:units}{1.7\MJysr}

\setsymbol{HFI:FIRAS:zero:point:accuracy:100GHz}{0.8}
\setsymbol{HFI:FIRAS:zero:point:accuracy:143GHz}{}
\setsymbol{HFI:FIRAS:zero:point:accuracy:217GHz}{}
\setsymbol{HFI:FIRAS:zero:point:accuracy:353GHz}{1.4}
\setsymbol{HFI:FIRAS:zero:point:accuracy:545GHz}{2.2}
\setsymbol{HFI:FIRAS:zero:point:accuracy:857GHz}{1.7}

% HFI diffuse source sensitivity unit conversion

\setsymbol{HFI:unit:conversion:100GHz:units}{0.2415\MJysrmK}
\setsymbol{HFI:unit:conversion:143GHz:units}{0.3694\MJysrmK}
\setsymbol{HFI:unit:conversion:217GHz:units}{0.4811\MJysrmK}
\setsymbol{HFI:unit:conversion:353GHz:units}{0.2883\MJysrmK}
\setsymbol{HFI:unit:conversion:545GHz:units}{0.05826\MJysrmK}
\setsymbol{HFI:unit:conversion:857GHz:units}{0.002238\MJysrmK}

\setsymbol{HFI:unit:conversion:100GHz}{0.2415}
\setsymbol{HFI:unit:conversion:143GHz}{0.3694}
\setsymbol{HFI:unit:conversion:217GHz}{0.4811}
\setsymbol{HFI:unit:conversion:353GHz}{0.2883}
\setsymbol{HFI:unit:conversion:545GHz}{0.05826}
\setsymbol{HFI:unit:conversion:857GHz}{0.002238}

% HFI Colour Correction for \alpha = -2, for V1.01 of the spectral bands

\setsymbol{HFI:colour:correction:alpha=-2:V1.01:100GHz}{0.9893}
\setsymbol{HFI:colour:correction:alpha=-2:V1.01:143GHz}{0.9759}
\setsymbol{HFI:colour:correction:alpha=-2:V1.01:217GHz}{1.0007}
\setsymbol{HFI:colour:correction:alpha=-2:V1.01:353GHz}{1.0028}
\setsymbol{HFI:colour:correction:alpha=-2:V1.01:545GHz}{1.0019}
\setsymbol{HFI:colour:correction:alpha=-2:V1.01:857GHz}{0.9889}

% HFI Colour Correction for \alpha = 0, for V1.01 of the spectral bands

\setsymbol{HFI:colour:correction:alpha=0:V1.01:100GHz}{1.0008}
\setsymbol{HFI:colour:correction:alpha=0:V1.01:143GHz}{1.0148}
\setsymbol{HFI:colour:correction:alpha=0:V1.01:217GHz}{0.9909}
\setsymbol{HFI:colour:correction:alpha=0:V1.01:353GHz}{0.9888}
\setsymbol{HFI:colour:correction:alpha=0:V1.01:545GHz}{0.9878}
\setsymbol{HFI:colour:correction:alpha=0:V1.01:857GHz}{1.0014}